\documentclass[
preprintnumbers,
twocolumn,
nofootinbib,
amsmath,amssymb,
 aps,
]{revtex4-2}

\usepackage{graphicx}
\usepackage{dcolumn}
\usepackage{bm}
\usepackage{hyperref}

\usepackage{comment}
\usepackage{caption}
\usepackage{subcaption}
\usepackage{booktabs}
\numberwithin{equation}{section}
\usepackage{color}
\usepackage{mathtools}
\usepackage[normalem]{ulem} 
\usepackage{amsmath}
\usepackage{orcidlink}

\usepackage{url}
\hypersetup{colorlinks=true, citecolor=cyan}

\newcommand{\be}{\begin{equation}}
\newcommand{\ee}{\end{equation}}

\definecolor{mygreen}{RGB}{0,130,0}

\begin{document}
\preprint{RESCEU-13/24}

\author{Daiki Watarai
\orcidlink{0009-0002-7569-5823}}
\affiliation{Graduate School of Science, The University of Tokyo, Tokyo 113-0033, Japan}
\affiliation{Research Center for the Early Universe (RESCEU), Graduate School of Science, The University of Tokyo, Tokyo 113-0033, Japan}


\newcommand*{\diff}{\,\mathrm{d}}

\date{\today}

\begin{abstract}
In a binary merger with a small mass ratio, as the secondary body approaches the innermost stable circular orbit (ISCO) of the primary black hole, the motion transitions from the adiabatic inspiral to the plunge governed by the geodesic equation. The plunge orbit is expected to excite the ringdown gravitational wave, which encodes information about the primary black hole's geometry. The details of the transition regime depend on the binary's mass ratio through radiation fluxes, which in turn influence the initial conditions for the plunge. As such, the mass ratio affects the post-ISCO ringdown gravitational wave excitation. In this study, we numerically investigate the mass ratio dependence of higher harmonic quasi-normal mode excitations in the post-ISCO gravitational waves of rapidly spinning black holes, based on the Teukolsky-Sasaki-Nakamura formalism. We consider the effect of mass ratio on the gravitational waves by accounting for the energy and angular momentum losses during the transition regime following the Ori-Thorne procedure. We examine two mass ratio scenarios: the intermediate mass ratio (IMR) and the extreme mass ratio (EMR). Our main finding is that higher harmonic quasinormal modes are significantly excited in an IMR merger involving a highly spinning primary black hole. This implies that detecting an IMR merger involving such a primary black hole with space-based gravitational wave interferometers can provide valuable opportunities to infer black hole properties or test general relativity with excellent precision.
\end{abstract}

\title{Ringdown of a postinnermost stable circular orbit of a rapidly spinning black hole: \\Mass ratio dependence of higher harmonic quasinormal mode excitation}
\maketitle

\section{Introduction}
Recent electromagnetic observations have revealed the presence of supermassive black holes (SMBHs) at the centers of galaxies, with masses ranging from $10^{5}$ to $10^{10}\:M_\odot$ (e.g.,~\cite{Greene_2010, John+_2013, Reynolds_2021}). Several studies suggest that SMBHs with masses around $10^{7}\:M_\odot$ tend to have rapid spins, including the near-extremal ones, as observed in both local SMBHs (e.g.,~\cite{Reynolds_2021}) and high-redshift analogs~\cite{Inayoshi2024}. Furthermore, investigations into the spin evolution of SMBHs through cosmological simulations indicate that SMBHs with masses around $10^{7}\:M_\odot$ are likely to possess the near-extremal spins~\cite{Dubois14,Sebastian_Springel_2019}.

A binary merger involving an SMBH is one of the main targets of space-based gravitational wave interferometers, such as the Laser Interferometer Space Antenna (LISA)~\cite{Amaro-Seoane_2017, Amaro-Seoane_2023}, which is planned to be launched in the next decade. LISA is sensitive to mHz gravitational waves (GWs) corresponding to the ringdown frequencies of BHs with masses around $10^7M_\odot$, which are expected to rotate rapidly from the observations and the simulation studies.

Ringdown GWs can be well modeled by a superposition of quasinormal modes (QNMs), which are fully determined by the BH's mass $M$ and spin $J$ in general relativity (GR) (e.g.,~\cite{Berti:2009kk, Konoplya:2011qq}). QNM frequencies $\omega = \omega_{lmn}(M, J)$ are labeled by the harmonic mode $l$, $m$, and the overtone number $n$. Accurate analyses of $\omega_{\ell mn}$ have shown that most QNMs with $\ell=m$ follow an asymptotic relation in the near-extremal spin limit: $2M\omega_{\ell mn} \rightarrow m$~\cite{Onozawa:1996ux, Cook:2014cta}. This relation indicates that QNMs are long-lived, but the character of the overtone tends to be lost at this limit. In other words, even if the signal-to-noise ratio is high, it is, in principle, difficult to identify each overtone with the same harmonic label\footnote{On the other hand, Ref.~\cite{Baibhav:2023clw} shows a general difficulty in extracting overtones in a practical sense.}. This fact suggests that we should focus on different harmonic QNMs since $\omega_{\ell mn}$ with different $\ell(=m)$ have clear separations, as the relation shows. Furthermore, the ringdown frequency of most QNMs increases monotonically as the BH rotates faster, resulting in more oscillations in the GW signal. As is well known in inspiral GW data analysis, more orbital cycles in the observational band enable us to measure the parameters more precisely\footnote{This is because the matched filtering method, which is generally adopted in GW data analysis, is sensitive to the GW phase evolution over time.}~\cite{Cutler:1994ys, Poisson:1995ef}. Analogously, the confident detection of higher harmonic QNMs will lead to a more precise estimation of the BH properties if they are sufficiently excited.

Relevant to these facts, Refs.~\cite{Oshita:2022pkc, watarai2024slowlydecayingringdownrapidly} investigate the observational forecasts with LISA, assuming the detection of GWs induced by an object falling radially into a rapidly spinning SMBH from infinity. First, Ref.~\cite{Oshita:2022pkc} elucidates the feasibility of BH spectroscopy with multiple QNMs and concludes that a highly spinning SMBH is advantageous since multiple QNMs, including higher harmonic ones, are efficiently excited. Furthermore, in Ref.~\cite{watarai2024slowlydecayingringdownrapidly}, it is quantitatively shown that precise measurements of a rapidly spinning SMBH's mass and spin are expected due to the higher harmonic QNMs being sufficiently excited to be informative. Importantly, the excitation of higher harmonic QNMs plays a crucial role in both cases. However, the event rate of such a plunging phenomenon is still unknown, and the effect of the mass ratio on the plunge waveforms has not been considered.

Apart from that, although there are still many discussions regarding the estimates of event rates (e.g.,~\cite{Amaro-Seoane_2023}), intermediate mass ratio (IMR) and extreme mass ratio (EMR) mergers are considered promising targets for space-based GW interferometers. The evolution of such a system is commonly divided into three phases: the inspiral, the transition, and the plunge. During the inspiral phase, the orbit of the secondary body evolves adiabatically due to gravitational radiation reaction, becoming quasicircular as the eccentricity decreases in the Keplerian regime~\cite{Peters:1964zz}. As the secondary approaches the innermost stable circular orbit (ISCO) of the primary BH, its motion gradually deviates from the adiabatic inspiral and enters the transition phase. In this regime, the motion is governed not only by radiation reaction but also by gravitational attraction~\cite{Ori:2000zn, Buonanno:2000ef}. Finally, the motion during the plunge phase is well described by geodesic motion since the radiation reaction is negligible. Ringdown GWs are expected to be excited at this stage.

The focus of this study is to evaluate the higher harmonic QNM excitations in the post-ISCO GWs and investigate the mass ratio dependence of these excitations. Since the details of the transition regime depend on the mass ratio of a binary through radiation fluxes, the initial conditions for the plunge will be affected by the mass ratio. We consider two mass ratio cases: the intermediate mass ratio (IMR) and the extreme mass ratio (EMR). Although numerical calculations for more generic plunge orbits can be found in Refs.~\cite{Apte:2019txp, Hughes:2019zmt, Lim:2019xrb}, we restrict our analysis to equatorial orbits since QNMs with $\ell=m$ are expected to be excited (for instance, see Fig.~1 of Ref.~\cite{Hughes:2019zmt}). 
Furthermore, Refs.~\cite{Hadar:2014dpa, Hadar:2015xpa, Hadar:2016vmk, Compere:2017hsi} show analytic expressions of the post-ISCO ringdown GW mode amplitudes for near extremally spinning primary BHs utilizing the near-horizon extremal Kerr (NHEK) limit. In this study, we numerically evaluate the amplitudes within the same computational framework for all BH spin cases, including the moderate spins.

This paper is organized as follows. In Sec.~\ref{sec:particle_motion}, we first briefly recap the test particle's motion in the Kerr spacetime and the Ori-Thorne procedure, a simple treatment of the transition regime, to derive the initial condition of the plunge. In Sec.~\ref{sec:post-isco_GWs}, we outline the Teukolsky-Sasaki-Nakamura formalism and give the details of the post-ISCO GW calculations by the Green function method. The numerical waveforms are shown in Sec.~\ref{sec:num_GW}. In Sec.~\ref{sec:higher_harmonic_QNM_excitations}, we assess higher harmonic QNM excitations by evaluating the relative amplitudes of higher harmonic modes to $(\ell, m)=(2,2)$ mode, which is the most dominant mode in the comparable mass ratio mergers. Finally, the discussion and conclusion are presented in Sec.~\ref{sec:discussion} and Sec.~\ref{sec:summary_and_conclusion}, respectively. We adopt the geometrical unit, $c=G=1$, throughout this paper.

\section{Post-innermost stable circular orbit plunging motion}
\label{sec:particle_motion}
We begin by confirming the convention used in this paper:
\begin{align*}
    & M : \mathrm{primary\:BH's\:mass}\:, \\
    & J :
    \mathrm{primary\:BH's\:spin}\:, \\
    & a :=J/M :\mathrm{primary\:BH's\:spin\:parameter}\:, \\
    & j := J/M^2 : \mathrm{dimensionless\:primary\:BH's\:spin\:parameter}\:, \\
    & \mu : \mathrm{secondary\:body's\:mass}\:, \\
    & q := \mu/M :\mathrm{mass\:ratio\:of\:a\:binary}\:,\\
    & \hat{E} : \mathrm{energy\:of\:the\:secondary\:body\:per\:its\:unit\:mass}\:, \\
    & \hat{L} : \begin{aligned}[t]
        &\mathrm{orbital\:angular\:momentum\:of\:the\:secondary\:body}\\
        &\mathrm{per\:its\:unit\:mass}\:.
    \end{aligned}
\end{align*}
Note that $\hat{E}$ and $\hat{L}$ are conserved for a geodesic motion. In this study, we set $M = 1$.

\subsection{Mass ratio and BH spin considered in the study}
\label{sec:cases}
We focus on the following values for the mass ratio and BH spin
\begin{itemize}
    \item $q=10^{-3}$ and $10^{-5}$\:.
    \item $j=0.8, 0.9, 0.99,$ and $0.999$\:.
\end{itemize}
The mass ratios $q=10^{-3}$ and $10^{-5}$ typically correspond to the intermediate mass ratio (IMR) and extreme mass ratio (EMR) cases, respectively. For the BH spin, we consider $j \geq 0.8$ and include the near-extremal cases $j=0.99$ and $0.999$. The BH spin of $j=0.999$ is close to the Thorne limit $j=0.998$, which is often assumed to be the maximum spin of an astrophysical BH~\cite{1974ApJ...191..507T}.

\subsection{Test particle's motion in the Kerr geometry}
\label{sec:motion_kerr}
Using Boyer-Lindquist coordinates, the background geometry is described by the Kerr metric with $M$ and $a$:
\begin{equation}
\begin{split}
    \diff s^2 =  & -\left(1-\frac{2Mr}{\Sigma}\right)\diff t^2 - \frac{4Mar\sin^2{\theta}}{\Sigma}\diff t \diff \phi + \frac{\Sigma}{\Delta} \diff r^2  \\ 
    &+ \Sigma \diff \theta^2 + \left( r^2 + a^2 + \frac{2Ma^2 r}{\Sigma}\sin^2{\theta} \right)\sin^2{\theta}\diff \phi^2\:,
\end{split} 
\end{equation}
where
\begin{align}
    &{\Delta(r) = r^2 - 2Mr + a^2} =: (r - r_+)(r - r_-)\:, \\
    &{\Sigma(r) = r^2 + a^2 \cos^2{\theta}\:.}
\end{align}
with 
\begin{equation}
    r_\pm \coloneqq M \pm \sqrt{M^2 + a^2}\:.
\end{equation}
For equatorial orbits ($\theta=\pi/2$), the equations of motion for a test particle are written as follows:
\begin{align}
    \label{eq:EOM_t}
    &r^2 \frac{\diff t}{\diff \tau} = -a(a\hat{E} - \hat{L}) + \frac{r^2 + a^2}{\Delta}P\:,\\
    \label{eq:EOM_r}
    &r^2 \frac{\diff r}{\diff \tau} = \pm\sqrt{Q}\:,\\
    \label{eq:EOM_phi}
    &r^2 \frac{\diff \phi}{\diff \tau} = -(a\hat{E} - \hat{L}) + \frac{a}{\Delta}P\:,
\end{align}
where 
\begin{gather}
    P(r) := \hat{E}(r^2 + a^2) - a\hat{L} \:, \\
    Q(r) := P^2 - \Delta \left[r^2 + (\hat{L} - a\hat{E})^2\right]\:, 
\end{gather}
and $\tau$ is the proper time of the particle (e.g.,~\cite{Misner:1973prb}). Circular motion is defined such that the orbit satisfies $\diff r/\diff \tau = \diff^2 r/\diff \tau^2 = 0$. The energy and angular momentum for circular orbits are then given by 
\begin{gather}
    \label{eq:E_hat}
    \hat{E} = \frac{1 - 2v^2 \pm jv^3}{\sqrt{1 - 3v^2 \pm 2jv^3}}\:, \\
    \label{eq:L_hat}
    \hat{L} = \frac{M(1 \mp 2jv^3 + j^2v^4)}{v\sqrt{ 1 - 3v^2 \pm 2jv^3 }}\:,
\end{gather}
where $v = (M/r)^{1/2}$, and $\pm$ denotes the prograde and retrograde orbits, respectively. Circular motion can exist if the denominator of Eq.~\eqref{eq:E_hat} is real.

The innermost stable circular orbit (ISCO) is defined by the orbit that minimizes $\hat{E}$. The radius denoted by $r_\mathrm{ISCO}$ is given by 
\begin{equation}
\label{eq:isco}
r_\mathrm{ISCO} = M \left[ 3 + Z_2 \mp \sqrt{(3-Z_1)(3+Z_1+2Z_2)} \right]\:,
\end{equation}
where
\begin{gather}
    Z_1 = 1+(1-j^2)^{1/3} [(1+j)^{1/3}+(1-j)^{1/3}]\:,\\
    Z_2 = \sqrt{ 3j^2 + Z_1^2 }\:.
\end{gather}
Inserting Eq.~\eqref{eq:isco} to Eqs.~\eqref{eq:E_hat} and \eqref{eq:L_hat}, we can derive energy and angular momentum at the ISCO $\hat{E}_\mathrm{ISCO}$ and $\hat{L}_\mathrm{ISCO}$.
The angular velocity at the ISCO $\Omega_\mathrm{ISCO} \coloneqq \diff \phi/\diff t (r=r_\mathrm{ISCO})$ is given by 
\begin{equation}
    \Omega_\mathrm{ISCO} = \pm \frac{M^{1/2}}{r_\mathrm{ISCO}^{3/2}\pm aM^{1/2}}\:.
\end{equation}
Similarly, the quantities evaluated at the ISCO are marked with ISCO in the following. 
Another relevant orbit to our study is the photon sphere (often referred to as light ring (LR) in the literature), which is given by
\begin{equation}
    r_\mathrm{LR} = 2M \left\{ 1 + \cos{\left[ \frac{2}{3}\cos^{-1}(\mp j) \right]} \right\}\:.
\end{equation}
Though more studies are needed for the association between ringdown GW excitation and the LR~\cite{Baibhav:2023clw}, we adopt the time at which the secondary object passes $r=r_\mathrm{LR}$ as a reference in Sec.~\ref{sec:evaluation_qnms}.

We consider the prograde orbits throughout this study, since QNMs with $\ell = m$ are expected to be significantly excited. Therefore, the upper sign is considered.

\subsection{The Ori-Thorne procedure}
\label{sec:OT}
As the secondary body approaches the ISCO due to radiation reaction, the orbital evolution ceases to be adiabatic and transitions smoothly from inspiral to plunge. This intermediate stage is referred to as the \textit{transition regime}, where radiation reaction and gravitational attraction comparably influence the orbital evolution (e.g.,~\cite{Ori:2000zn, Buonanno:2000ef, Kesden:2011ma, Compere:2019cqe, Burke:2019yek}). Since the radiation fluxes depend on the mass ratio $q$, the duration and width of the transition regime are also $q$-dependent. This dependence affects the initial conditions for the plunge. Notably, these changes significantly impact the excitation of higher harmonic QNMs, as we will discuss in Sec.~\ref{sec:evaluation_qnms}.

To determine the initial conditions for the plunge, we adopt the simplest approach to the transition regime proposed by Ori and Thorne~\cite{Ori:2000zn} (hereafter OT). Their basic idea is to include a gradual change in the potential in the radial geodesic equation due to radiative losses. The radiative rates are assumed to be those at the ISCO. Following their convention, we define quantities normalized by $M$ with a tilde, such as $\tilde{E} = \hat{E}$, $\tilde{L} = \hat{L}/M$, $\tilde{t} = t/M$, $\tilde{r} = r/M$, $\tilde{\Omega} = M\Omega$, and so on. We start from the transformed radial equation of motion (EOM) [Eq.~\eqref{eq:EOM_r}] at the ISCO:
\begin{equation}
    \label{eq:dr_dtau}
    \left(\frac{\diff \tilde{r}}{\diff \tilde {\tau}}\right)^2 = \tilde{E}^2_\mathrm{ISCO} - V(\tilde{r}_\mathrm{ISCO}, \tilde{E}_\mathrm{ISCO}, \tilde{L}_\mathrm{ISCO})\:, 
\end{equation}
where $V$ is the effective potential, given by
\begin{equation}
    \label{V_eff}
    \begin{split}
        V(\tilde{r}, \tilde{E}, \tilde{L}) = &\tilde{E}^2 - \frac{1}{\tilde{r}^4} \left\{ \left[\tilde{E}(\tilde{r}^2 + j^2) - \tilde{L}j\right]^2 \right. \\ 
        &\left. - (\tilde{r}^2 - 2\tilde{r} + j^2)\left[\tilde{r}^2 + (\tilde{L} - \tilde{E}j)^2 \right] \right\}\:.
    \end{split}
\end{equation}
Taking the derivative of Eq.~\eqref{eq:dr_dtau} with respect to $\tilde{\tau}$ and using the energy radiation formula,
\begin{equation}
    \label{eq:dE_dt}
    \frac{\diff E}{\diff t} = -\frac{32}{5} q^2 \tilde{\Omega}^{10/3} \dot{\mathcal{E}}\:,
\end{equation}
where $E = \mu\hat{E}$, and $\dot{\mathcal{E}}$ is the general relativistic correction to the Newtonian quadrupole radiation\footnote{The value of $\dot{\mathcal{E}}$ can be found in Table~I of OT (originally, Table~II of Ref.~\cite{Finn:2000sy}).
 Each contains contributions from a sufficient number of multipole modes to an accuracy of four significant digits.},
we reach
\begin{equation}
\begin{split}
    \label{eq:d2r_dtau2}
    \frac{\diff^2 \tilde{r}}{\diff \tilde{\tau}^2} = -\frac{1}{2}\frac{\partial V}{\partial \tilde{r}} + \mathcal{O}(q)\:.
\end{split}
\end{equation}
The term $\mathcal{O}(q)$ includes the self-force, which we neglect here as it is subdominant when $q$ is sufficiently small. We assume that throughout the transition regime, as well as during the adiabatic inspiral, the secondary mass remains in a nearly circular orbit. Specifically, the change in radius $\Delta r$ over a period is much smaller than the radius $r$ itself, i.e., $\Delta r \ll r$, and the radiation fluxes remain the same as those at the ISCO. These mean that
\begin{equation}
    \label{eq:dE_omegadL}
    \frac{\diff \tilde{E}}{\diff \tilde{\tau}}\bigg|_\mathrm{ISCO} = \tilde{\Omega}_\mathrm{ISCO} \frac{\diff \tilde{L}}{\diff \tilde{\tau}}\bigg|_\mathrm{ISCO}\:,
\end{equation}
holds throughout the entire transition regime. Using Eq.~\eqref{eq:dE_omegadL} and introducing small parameters from the ISCO values as $\xi = \tilde{L}-\tilde{L}_\mathrm{ISCO}$ and $R = \tilde{r}-\tilde{r}_\mathrm{ISCO}$, the potential $V$ can be Taylor expanded as
\begin{equation}
    \label{eq:potential}
    V(R, \xi) \simeq \frac{2\alpha}{3}R^3 - 2\beta R\xi + \mathrm{const.}\:,
\end{equation}
up to cubic terms in $R$ and linear terms in $\xi$ (needed to be kept for calculation of the motion), where
\begin{gather}
    \label{eq:alpha}
    \alpha = \frac{1}{4}\left[ \frac{\partial^3 V(\tilde{r}, \tilde{E}, \tilde{L})}{\partial\tilde{r}^3} \right]_\mathrm{ISCO}\:, \\
    \label{eq:beta}
    \beta = -\frac{1}{2}\left[ \frac{\partial^2 V(\tilde{r}, \tilde{E}, \tilde{L})}{\partial\tilde{L}\partial\tilde{r}} + \tilde{\Omega} \frac{\partial^2 V(\tilde{r}, \tilde{E}, \tilde{L})}{\partial \tilde{E} \partial \tilde{r}} \right]_\mathrm{ISCO}\:.
\end{gather}
Inserting Eqs.~\eqref{eq:potential}, \eqref{eq:alpha}, and \eqref{eq:beta} into Eq.~\eqref{eq:d2r_dtau2}, EOM can be written as
\begin{equation}
    \label{eq:d2R_dtau2}
    \frac{\diff^2 R}{\diff \tilde{\tau}^2} = -\alpha R^2 + \beta \xi = -\alpha R^2 - q\beta \kappa \tilde{\tau}\:,
\end{equation}
where 
\begin{equation}
    \kappa = \frac{32}{5}{\tilde{\Omega}_\mathrm{ISCO}}^{7/3} \frac{1+j/\tilde{r}^{3/2}_\mathrm{ISCO}}{\sqrt{1-3/\tilde{r}_\mathrm{ISCO}+2j/\tilde{r}_\mathrm{ISCO}^{3/2}}} \dot{\mathcal{E}}\:.
\end{equation}
Finally, we convert Eq.~\eqref{eq:d2R_dtau2} into a dimensionless form,
\begin{equation}
    \label{eq:EOM_transition}
    \frac{\diff^2 X}{\diff T^2} = -X^2 - T\:,
\end{equation}
with 
\begin{equation}
    R =: q^{2/5}R_o X\:, \:\:\:\tilde{\tau} =: q^{-1/5}\tau_o T\:,
\end{equation}
where
\begin{equation}
    R_o := (\beta \kappa)^{2/5}\alpha^{-3/5}\:,\:\:\:\tau_o := (\alpha\beta\kappa)^{-1/5}\:.
\end{equation}
Since the motion smoothly connects with the plunge orbit as $X \rightarrow -\infty$ at late times, we can neglect the second term in Eq.~\eqref{eq:EOM_transition} and derive the approximate solution:
\begin{equation}
    X = \frac{-6}{(T_\mathrm{plunge}-T)^2}\:,\:\:\:\:T\rightarrow T_\mathrm{plunge}\:,
\end{equation}
where $T_\mathrm{plunge}=3.412$ numerically, and the solution diverges at $T=T_\mathrm{plunge}$.

\begin{table*}[t]
\begin{ruledtabular}
\caption{\label{tb:fractional_errors_from_isco_values}%
Radiated energy, angular momentum, and radial width during the transition regime. Each value is normalized by its corresponding value at the ISCO.
}
\begin{tabular}{cccccc}
Mass ratio & primary BH spin &  $\Delta \tilde{E}/\tilde{E}_\mathrm{ISCO}(\propto q^{4/5})\:[\%]$ & $\Delta \tilde{L}/\tilde{L}_\mathrm{ISCO}(\propto q^{4/5})\:[\%]$ & $\Delta \tilde{r}/\tilde{r}_\mathrm{ISCO}(\propto q^{2/5})$\:[\%]& \\
\colrule
$q=10^{-3}$ & $j=0.8$ &  $0.316$ & $0.671$ & $6.60$ & \\
${}$ & $j=0.9$ & $0.576$ & $1.03$ & $7.61$ & \\
${}$ & $j=0.99$ & $1.92$  & $2.46$ & $7.65$ & \\
${}$ & $j=0.999$ & $3.53$  & $3.97$ & $5.67$ & \\
\colrule
$q=10^{-5}$ & $j=0.8$ &  $0.00794$ & $0.0169$ & $0.417$ & \\
${}$ & $j=0.9$ & $0.0145$ & $0.0258$ & $0.480$ & \\
${}$ & $j=0.99$ & $0.0483$ & $0.0622$ & $0.483$ & \\
${}$ & $j=0.999$ & $0.0886$  & $0.0996$ & $0.358$ & \\
\end{tabular}
\end{ruledtabular}
\end{table*}

\begin{figure*}[t]
\includegraphics[scale=0.35]{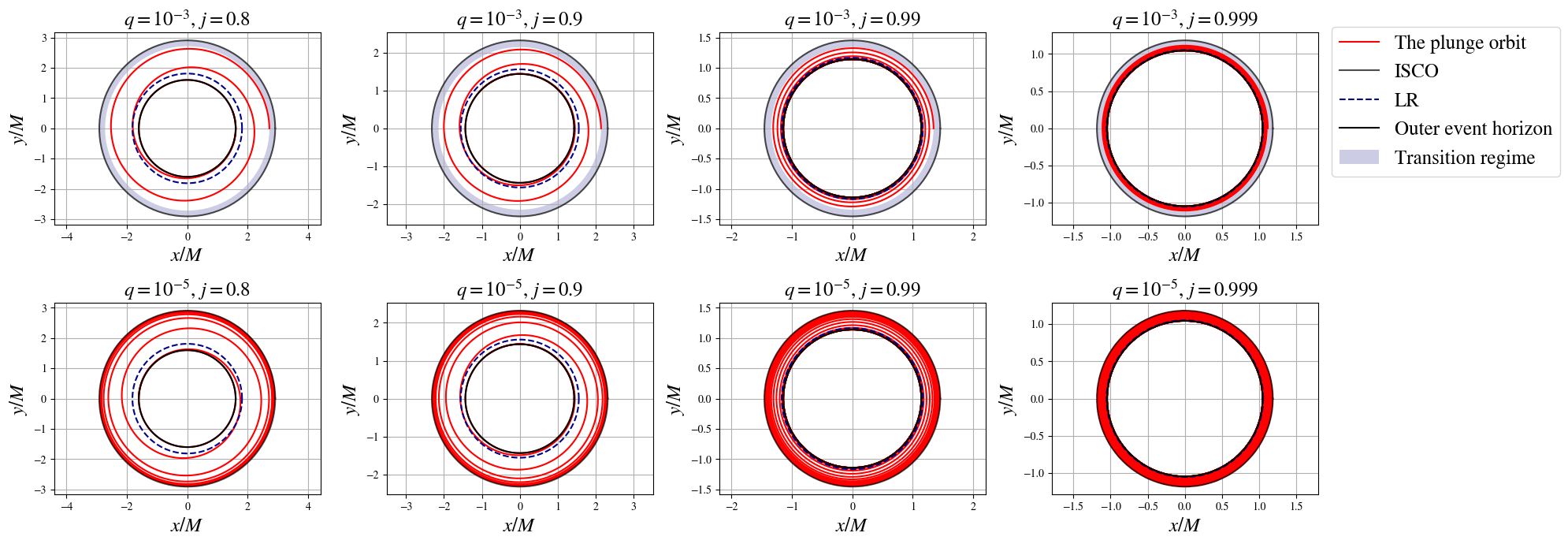}
\caption{\label{fig:Particle_trajectory}
The plunge orbit after the transition regime with mass ratios of $q=10^{-3}$ (top panels) and $q=10^{-5}$ (bottom panels) for primary BH spins of $j=0.8$, $0.9$, $0.99$, and $0.999$ (from left to right). In each panel, the red line shows the plunge orbit, which is a geodesic with energy $\hat{E}_\mathrm{plunge} = \hat{E}_\mathrm{ISCO} - \Delta \hat{E}$ and angular momentum $\hat{L}_\mathrm{plunge} = \hat{L}_\mathrm{ISCO} - \Delta \hat{L}$, starting from $r = r_\mathrm{ISCO} - \Delta r$. The gray solid, navy dashed, and navy solid lines represent the ISCO, LR, and outer event horizon, respectively. The light blue region marks the transition regime for each case.
}
\end{figure*}

\subsection{Initial condition and trajectory of the plunge orbit}
\label{sec:plunge_orbit}
Assuming that the radiation fluxes remain the same as those at the ISCO, the energy and angular momentum losses in the transition regime can be approximated by
\begin{align}
    \label{eq:Delta_L}
    \Delta \tilde{L} &= (\kappa \tau_0 T_\mathrm{plunge}) q^{4/5}\:, \\
    \label{eq:Delta_E}
    \Delta \tilde{E} &= \tilde{\Omega}_\mathrm{ISCO} (\kappa \tau_0 T_\mathrm{plunge}) q^{4/5}\:.
\end{align}
Combining Eq.~\eqref{eq:Delta_L} with Eqs.~\eqref{eq:dr_dtau}, \eqref{eq:dE_dt}, and $\diff \tilde{\tau}/\diff \tilde{t} = \sqrt{1-3/\tilde{r}+2j}$, $\Delta \tilde{r}$ is given by
\begin{equation}
\label{eq:Delta_tilde_r}
    \Delta \tilde{r} = \frac{\Delta \tilde{E}}{\kappa q \tilde{\Omega}_\mathrm{ISCO}} \frac{\diff \tilde{r}}{\diff \tilde {\tau}}\bigg|_\mathrm{ISCO} \propto q^{2/5}\:.
\end{equation}
Thus, we start the plunge orbit at $\tilde{r} = \tilde{r}_\mathrm{ISCO} - \Delta \tilde{r}$ with 
\begin{align}
    \label{eq:tilde_E_plunge}
    \tilde{E}_\mathrm{plunge} = \tilde{E}_\mathrm{ISCO} - \Delta \tilde{E}\:, \\
    \label{eq:tilde_L_plunge}
    \tilde{L}_\mathrm{plunge} = \tilde{L}_\mathrm{ISCO} - \Delta \tilde{L}\:.
\end{align}
Equations~\eqref{eq:Delta_L}-\eqref{eq:Delta_tilde_r} indicate that $\Delta \tilde{L}, \Delta \tilde{E}$, and $\Delta \tilde{r}$ are sensitive to the mass ratio $q$ because the leading mass ratio effects appear at the order of $q^{4/5}$ or $q^{2/5}$.
Table~\ref{tb:fractional_errors_from_isco_values} shows $\Delta \tilde{E}/\tilde{E}_\mathrm{ISCO}$, $\Delta \tilde{L}/\tilde{L}_\mathrm{ISCO}$, and $\Delta \tilde{r}/\tilde{r}_\mathrm{ISCO}$ for $j=0.8$, $0.9$, $0.99$, and $0.999$ with $q=10^{-3}$ and $10^{-5}$. $\Delta \tilde{E}/\tilde{E}_\mathrm{ISCO}$ and $\Delta \tilde{L}/\tilde{L}_\mathrm{ISCO}$ have larger values as $q$ and $j$ increase, which is mainly because gravitational radiation during the plunge becomes more significant for larger $q$ and $j$. This trend holds for $\Delta \tilde{r}/\tilde{r}_\mathrm{ISCO}$, except in the case of $j=0.999$, where $r_\mathrm{ISCO} - r_+$ is small.

Figure~\ref{fig:Particle_trajectory} shows the plunge trajectories with mass ratio of $q=10^{-3}$ (top panels) and $q=10^{-5}$ (bottom panels) for $j=0.8, 0.9, 0.99$ and $0.999$ (from left to right). A red line in each panel depicts the plunge orbit which is a geodesic with the energy $\hat{E}_\mathrm{plunge}$ and angular momentum $\hat{L}_\mathrm{plunge}$ starting from $r=r_\mathrm{ISCO}-\Delta r$. In each panel, gray solid, navy dashed, and navy solid lines show the ISCO, LR, and outer event horizon, respectively. The light blue region marks the transition regime for each case. The panels visually show the transition regime has a wider radial width in IMR cases and, more cycles can be seen due to stronger frame-dragging effects as $j$ increases.

\subsection{Remark}
It should be noted that, in addition to OT, several studies propose more general treatments of the transition regime (e.g.,~\cite{Kesden:2011ma, Compere:2019cqe, Apte:2019txp, K_chler_2024}). For instance, Ref.~\cite{Kesden:2011ma} relaxes Eq.~\eqref{eq:dE_omegadL} and addresses an inconsistency in the normalization of the secondary object's four-velocity that arises in the OT procedure. In addition, Ref.~\cite{Compere:2019cqe} considers the near-extremally spinning case, where another small parameter $\lambda:=\sqrt{1-j^2}$ contributes to the equation of motion in addition to mass ratio $q$. Although these references will provide more appropriate treatments for the motion in the transition regime, we adopt the OT procedure as the most simplified treatment of the transition regime. Possible changes in the final result by considering the generic prescriptions are discussed in Sec.~\ref{sec:discussion}.

\section{Gravitational waves from the plunge orbit}
\label{sec:post-isco_GWs}
We numerically compute the post-ISCO GWs induced by the plunge orbit using the Teukolsky-Sasaki-Nakamura formalism~\cite{Teukolsky:1973ha, Press:1973zz, Teukolsky:1974yv, Sasaki:1981sx} (see Refs.~\cite{Hadar:2009ip, Hadar:2011vj} for the Schwarzschild case). First, we briefly outline the formalism for calculating the GWs in Sec.~\ref{sec:TSN}. Unlike previous studies (e.g.,~\cite{Mino:2008at, Sundararajan:2010sr, Lim:2019xrb, Hughes:2019zmt}), we adopt the Sasaki-Nakamura (SN) formalism. The main reason is to address concerns about the divergent nature of the source term in the Teukolsky equation, even though the compact-support sources we consider are generally not relevant to this issue. 
We thus derive the explicit form of the source term for a generic geodesic orbit in Appendix~\ref{app:SN_source}\footnote{The source terms for specific cases can be found in the literature (e.g.,~\cite{Sasaki:1981sx, Kojima:1984cj, Nakamura:1987zz, Shibata:1993yf}). References~\cite{Saijo:1996iz, Saijo:1998mn} provide source terms for a generic geodesic trajectory (especially, Ref.~\cite{Saijo:1998mn} presents one for a spinning particle), but our expression shown in Appendix~\ref{app:SN_source} differs due to a difference in performing partial integrals. Our expression is derived such that it reduces to that in Ref.~\cite{Kojima:1984cj} when taking $\hat{E}=1$.}. 

In Sec.~\ref{sec:num_GW}, we present the numerical gravitational waveforms. The details of the numerical calculation setup are provided in Appendix~\ref{app:comp_details}.

\subsection{Teukolsky-Sasaki-Nakamura formalism}
\label{sec:TSN}
Gravitational radiations at infinity are described by a Newman-Penrose quantity $\psi_4 = -C_\mathrm{\alpha\beta\gamma\delta} n^\alpha \overline{m}^\beta n^\gamma \overline{m}^\delta$, where $C_\mathrm{\alpha\beta\gamma\delta}$ is the Weyl tensor, and $n^\alpha$ and $m^\alpha$ are the tetrads defined by
\begin{align}
    &n^\alpha = \frac{1}{2\Sigma}\left( r^2+a^2, -\Delta, 0, a \right)\:, \\
    &m^\alpha = \frac{\overline{\rho}}{\sqrt{2}} \left( \mathrm{i}a\sin\theta, 0, 1, \frac{\mathrm{i}}{\sin\theta} \right)\:.
    \end{align}
Here,
\begin{equation}
    \rho(r, \theta) = \frac{1}{r-\mathrm{i}a\cos\theta}\:.
\end{equation}
$\psi_4$ can be decomposed in terms of the spin-weighted spheroidal harmonics ${}_{-2}S^{a\omega}_{\ell m}$ to
\begin{equation}
\label{eq:psi_4}
    \rho^{-4} \psi_4 = \sum_{l,m} \int \diff\omega\:\mathrm{e}^{-\mathrm{i}\omega t} R_{\ell m\omega}(r) {}_\mathrm{-2}S^{a\omega}_{\ell m}(\theta)\frac{\mathrm{e}^{\mathrm{i}m\phi}}{\sqrt{2\pi}}\:,
\end{equation}
and the gravitational waveform is given by
\begin{equation}
\label{eq:h_psi_4}
    h_+ - \mathrm{i}h_\times = -\frac{2}{\omega^2}\psi_4\:.
\end{equation}
The spin-weighted spheroidal harmonic ${}_\mathrm{-2}S^{a\omega}_{\ell m}(\theta)$ satisfies
\begin{equation}
\begin{split}
    \Bigg[ &\frac{1}{\sin\theta}\frac{\diff }{\diff \theta}\left( \sin^2\theta \frac{\diff}{\diff \theta} \right) - a^2\omega^2 \sin^2\theta - \frac{(m-2\cos\theta)}{\sin^2\theta}\\ &+ 4a\omega \cos\theta - 2 + 2ma\omega + \lambda \Bigg] {}_\mathrm{-2}S^{a\omega}_{\ell m}(\theta) = 0 \:,
\end{split}
\end{equation}
where $\lambda$ is the eigenvalue.
We fix the normalization such that
\begin{equation}
\label{eq:normalization_S_lm}
    \int^\pi_0 \big| {}_\mathrm{-2}S^{a\omega}_{\ell m}(\theta) \big|^2 \sin\theta \diff \theta=1\:,
\end{equation}
and implement it based on Ref.~\cite{Berti:2005gp}.
The radial function $R_{\ell m\omega}(r)$ satisfies the radial Teukolsky equation,
\begin{equation}
    \label{eq:radial_teukolsky}
    \Delta^2 \frac{\diff }{\diff r } \left( \frac{1}{\Delta} \frac{\diff R_{\ell m\omega}}{\diff r} \right) - V(r)R_{\ell m\omega} = T_{\ell m\omega}\:, 
\end{equation}
where $V(r)$ is given by
\begin{equation}
    V(r) = -\frac{K^2 + 4\mathrm{i}(r-M)K}{\Delta} + 8\mathrm{i}\omega r + \lambda\:.
\end{equation}
The source term $T_{\ell m\omega}$ is given in Ref.~\cite{Geroch:1973am} by
\begin{equation}
\label{eq:T_lmw}
    T_{\ell m\omega} = 4\int \diff\Omega \diff t\:\rho^{-5} \overline{\rho}^{-1} (B'_2+B'^*_2) \frac{\mathrm{e}^{-\mathrm{i}\omega t + \mathrm{i}m\phi}}{\sqrt{2\pi}}{}_\mathrm{-2}S^{a\omega}_{\ell m}(\theta)\:.
\end{equation}
Explicit forms of $B'_2$ and $B'^*_2$ are shown in Appendix~\ref{app:SN_source}. 

In our calculation, we adopt the Sasaki-Nakamura (SN) formalism rather than solving Eq.~\eqref{eq:radial_teukolsky}. 
In this formalism, we work with the SN variable $X_{\ell m\omega}$ which is related to $R_{\ell m\omega}$ by the relation
\begin{equation}
\label{eq:R_X}
\begin{split}
    R_{\ell m\omega} = \frac{1}{\gamma} \Bigg[ &\frac{\alpha \Delta + \beta'}{\sqrt{r^2+a^2}} X_{\ell m\omega}- \frac{\beta}{\Delta}\left(\frac{\Delta X_{\ell m\omega}}{\sqrt{r^2+a^2}}\right)' \Bigg]\:,    
\end{split}
\end{equation}
where 
\begin{align}
    \alpha &= 3\mathrm{i}K' + \lambda + \frac{6\Delta}{r^2} - \mathrm{i}\frac{K \beta}{\Delta^2}\:,\\
    \beta &= \Delta \left(-2\mathrm{i}K + \Delta' - \frac{4\Delta}{r}\right)\:,\\
    \gamma &= c_0 + \frac{c_1}{r} + \frac{c_2}{r^2} + \frac{c_3}{r^3} + \frac{c_4}{r^4} \:,    
\end{align}
with
\begin{align}
    \label{eq:c_0}
    c_0 &= -12\mathrm{i}\omega M + \lambda(\lambda+2) - 12a\omega (a\omega-m)\:, \\
    c_1 &= 8\mathrm{i}a[3a\omega - \lambda(a\omega - m)]\:,\\
    c_2 &= -24\mathrm{i}a M(a\omega-m) + 12a^2 [1-2(a\omega-m)^2]\:, \label{eq:c2} \\
    c_3 &= 24\mathrm{i} a^3 (a\omega-m) - 24Ma^2\:, \\
    c_4 &= 12a^4\:, 
\end{align} 
and the prime denotes the partial derivative with respect to $r$. $X_{\ell m\omega}$ follows the SN equation
\begin{equation}
\label{eq:SN_eq}
    \left[ \frac{\diff^2}{\diff {r_\ast}^2} - F_{\ell m}(r)\frac{\diff}{\diff r_\ast} - U_{\ell m}(r) \right] X_{\ell m\omega} (r_\ast) = \mathcal{S}_{\ell m\omega}(r)\:,
\end{equation}
where $r_\ast$ is the tortoise coordinate which we define by
\begin{equation}
    r_\ast = r + \frac{2M}{r_+-r_-}\left[ r_+\ln\left( \frac{r-r_+}{2M} \right) - r_-\ln\left( \frac{r-r_-}{2M} \right) \right]\:,
\end{equation}
and 
\begin{align}
    &F_{\ell m}(r) = \frac{\gamma'}{\gamma} \frac{\Delta}{r^2+a^2}\:,\\    
    &U_{\ell m}(r) = \frac{\Delta U_1}{(r^2+a^2)^2} + G^2 + \frac{\Delta G'}{r^2+a^2} - FG\:,
\end{align}
where
\begin{align}
    G &= -\frac{2(r-M)}{r^2+a^2} + \frac{r\Delta}{(r^2+a^2)^2}\:, \\
    U_1 &= V + \frac{\Delta^2}{\beta}\left[ \left( 2\alpha + \frac{\beta'}{\Delta} \right)' - \frac{\gamma'}{\gamma}\left( \alpha + \frac{\beta'}{\Delta} \right) \right]\:.
\end{align}
$\mathcal{S}_{\ell m\omega}$ is the source term whose form is expressed as
\begin{equation}
\label{eq:S_lnw}
    \mathcal{S}_{\ell m\omega} = \frac{\gamma \Delta}{(r^2+a^2)^{3/2}r^2} \mathcal{W}\:\mathrm{exp}\left[ - \int^r \frac{K}{\Delta} \diff r \right]\:.
\end{equation}
We derive the explicit form of $\mathcal{W}$ for a generic geodesic orbit in Appendix~\ref{app:SN_source}. This term contains all the details of the source's trajectory derived in Sec.~\ref{sec:particle_motion}.

\begin{figure*}[t]
\includegraphics[scale=0.48]{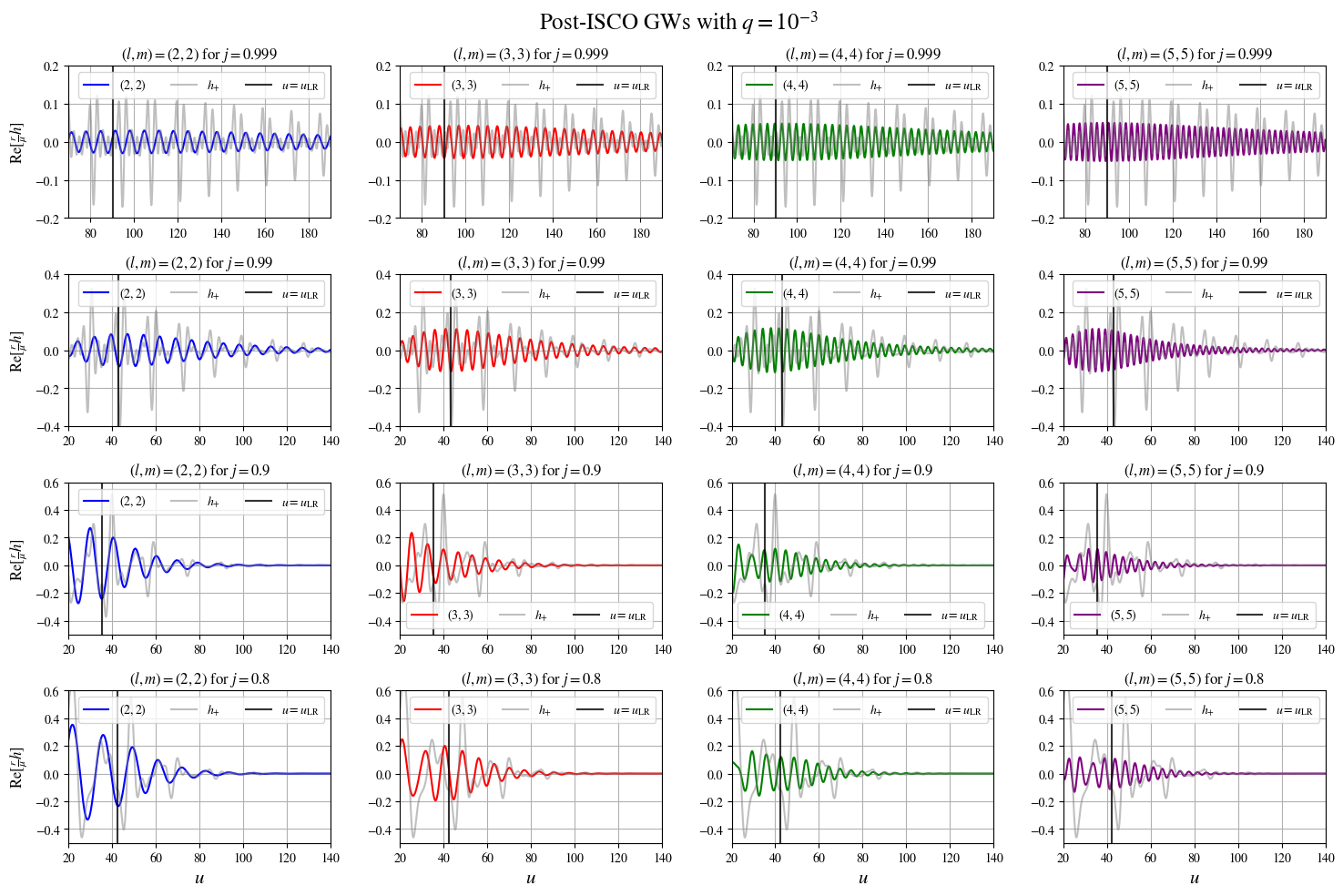}
\caption{\label{fig:Post-isco_GWs_q_3}
Numerical gravitational waveforms from the plunge orbit with $q=10^{-3}$ for primary BH spins of $j=0.08$, $0.9$, $0.99$, and $0.999$ (from bottom to top). The blue, red, green, and purple lines represent the real parts of $h_{22}$, $h_{33}$, $h_{44}$, and $h_{55}$, respectively. The gray and black lines show the total strain $h_+ = h_{22} + h_{33} + h_{44} + h_{55}$ and $u = u_\mathrm{LR}$, the retarded time at which the secondary object passes the LR.
}
\end{figure*}

\begin{figure*}[t]
\includegraphics[scale=0.48]{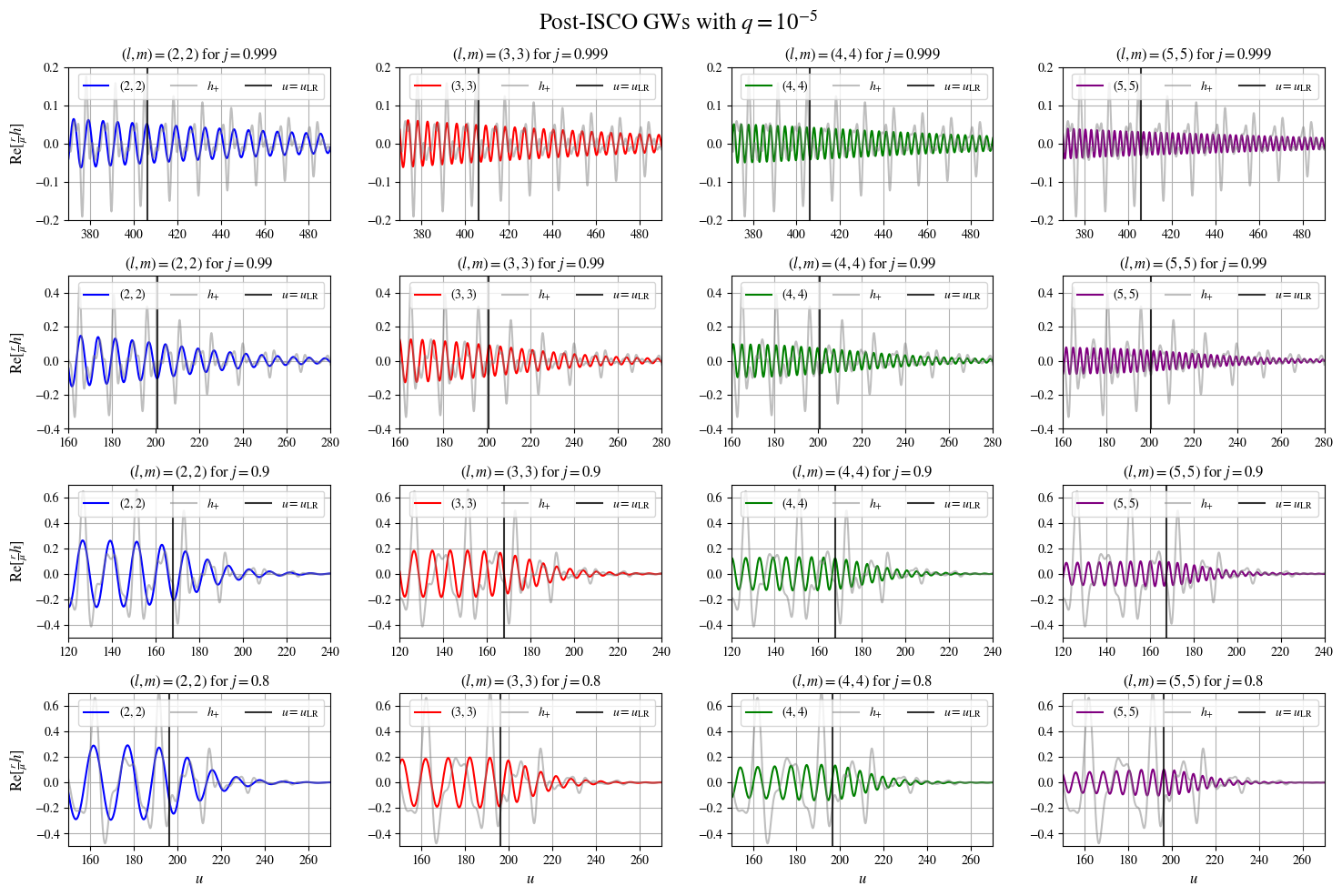}
\caption{\label{fig:Post-isco_GWs_q_5}
Same as Fig.~\ref{fig:Post-isco_GWs_q_3}, but the mass ratio of $q=10^{-5}$.
}
\end{figure*}

Equation~\eqref{eq:SN_eq} can be solved by using the Green function method. In order to do this, we prepare two independent homogeneous solutions to Eq.~\eqref{eq:SN_eq},
\begin{equation}
\label{eq:in_SN}
X^\mathrm{in}_{\ell m\omega}( r_\ast) \rightarrow \left\{
\begin{array}{ll}
A^\mathrm{trans}_{\ell m\omega}\mathrm{e}^{-\mathrm{i}kr_\ast}\:, & r_\ast \rightarrow  -\infty\\
A^\mathrm{ref}_{\ell m\omega}\mathrm{e}^{\mathrm{i}\omega r_\ast} + A^\mathrm{inc}_{\ell m\omega}\mathrm{e}^{-\mathrm{i}\omega r_\ast}\:, & r_\ast\rightarrow +\infty \\
\end{array}\;,
\right.
\end{equation}
\begin{equation}
\label{eq:out_SN}
X^\mathrm{out}_{\ell m\omega}( r_\ast) \rightarrow \left\{
\begin{array}{ll}
B^\mathrm{ref}_{\ell m\omega}\mathrm{e}^{-\mathrm{i}kr_\ast} + B^\mathrm{inc}_{\ell m\omega}\mathrm{e}^{\mathrm{i}kr_\ast}\:, & r_\ast \rightarrow  -\infty\\
B^\mathrm{trans}_{\ell m\omega}\mathrm{e}^{\mathrm{i}\omega r_\ast}\:, & r_\ast\rightarrow +\infty \\
\end{array}\;,
\right.
\end{equation}
where $k=\omega-ma/(2r_+)$.
Furthermore, we introduce a new variable $\xi$, which is defined by a relation $\diff \xi = \gamma \diff r_\ast$, and rewrite the inhomogeneous SN equation as
\begin{equation}
    \left(\frac{\diff^2}{\diff \xi^2} - \frac{U_{\ell m}}{\xi^2}\right) X_{\ell m\omega} = \frac{\mathcal{S}_{\ell m\omega}}{\xi^2}\:.
\end{equation}
Then, the solution can be constructed as
\begin{equation}
    X_{\ell m\omega}(\xi) = \int^\infty_{-\infty} \mathcal{G}(\xi,\xi')\mathcal{S}_{\ell m\omega}(\xi') \diff \xi'\:,
\end{equation}
where $\mathcal{G}(\xi,\xi')$ is the Green function given by
\begin{equation}
\begin{split}
    \mathcal{G}(\xi,\xi') = \frac{1}{W}\Big[ &X^\mathrm{in}_{\ell m\omega}(\xi) X^\mathrm{out}_{\ell m\omega}(\xi') \theta(\xi'-\xi)\\ &+ X^\mathrm{in}_{\ell m\omega}(\xi') X^\mathrm{out}_{\ell m\omega}(\xi) \theta(\xi-\xi') \Big]\:,
\end{split}
\end{equation}
with $W$ is the Wronskian defined by
\begin{equation}
    W = X^\mathrm{in}_{\ell m\omega}\frac{\diff X^\mathrm{out}_{\ell m\omega}}{\diff \xi} - X^\mathrm{out}_{\ell m\omega}\frac{\diff X^\mathrm{in}_{\ell m\omega}}{\diff \xi} = \frac{2\mathrm{i}\omega}{c_0}A^\mathrm{inc}_{\ell m\omega}B^\mathrm{trans}_{\ell m\omega}\:,
\end{equation}
and is constant in $\xi$.
$\mathcal{G}$ is constructed such that $X_{\ell m \omega}$ satisfies the physically appropriate boundary conditions: a purely ingoing wave at the horizon and a purely outgoing wave at infinity. The asymptotic solution at infinity is given by
\begin{equation}
    X_{\ell m\omega}(r_\ast \rightarrow \infty) = X^{(\infty)}_{\ell m\omega}\:\mathrm{e}^{\mathrm{i}\omega r_\ast}\:,
\end{equation}
where
\begin{equation}
\label{eq:X_infty}
    X^{(\infty)}_{\ell m\omega} = \frac{c_0}{2\mathrm{i}\omega A^\mathrm{inc}_{\ell m\omega}}\int^{\infty}_{-\infty} \frac{X^\mathrm{in}_{\ell m\omega}\mathcal{S}_{\ell m\omega}}{\gamma} \diff r_\ast\:.
\end{equation}
Using Eq.~\eqref{eq:R_X}, we can convert $X_{\ell m\omega}$ to $R_{\ell m\omega}$ at infinity as
\begin{equation}
\label{eq:R_to_X_infty}
    R_{\ell m\omega}(r\rightarrow \infty) =\:-\frac{4 \omega^2}{c_0} X^{(\infty)}_{\ell m\omega}\:r^3 \mathrm{e}^{\mathrm{i}\omega r_\ast}\:.
\end{equation}
Finally, using Eqs.~\eqref{eq:psi_4}, \eqref{eq:h_psi_4}, and \eqref{eq:R_to_X_infty}, the waveform is given by
\begin{equation}
\label{eq:strain}
\begin{split}
    h_+-\mathrm{i}h_\times &= \:\sum_{l,m}\:h_{\ell m}(t)\\
    &=\:\sum_{l,m} \int^\infty_{-\infty} \mathrm{e}^{-\mathrm{i}\omega u} \tilde{h}_{\ell m}(\omega) \diff \omega\:,    
\end{split}
\end{equation}
where $u$ is the retarded time defined by $u\coloneqq t-r_\ast$, and  
\begin{equation}
\label{eq:strain_lm}
    \tilde{h}_{\ell m}(\omega) = \frac{8}{r} \frac{X^{(\infty)}_{\ell m\omega}}{c_0} {}_\mathrm{-2}S^{a\omega}_{\ell m}(\iota)\frac{\mathrm{e}^{\mathrm{i}m\phi}}{\sqrt{2\pi}}\:,
\end{equation}
where $\iota$ is the inclination angle.
Note that $h_{\ell m}$ and $\tilde{h}_{\ell m}$ defined here include ${}_{-2}S^{a\omega}_{\ell m}$. In the following except in Sec.~\ref{sec:inclination}, we set $\iota$ to $\pi/2$, corresponding to the edge-on binary. The dependence of $\iota$ on the final results will be discussed in Sec.~\ref{sec:inclination}.
\begin{widetext}
To derive $X^\mathrm{in}_{\ell m \omega}$, we numerically integrate Eq.~\eqref{eq:SN_eq} from the horizon imposing the purely ingoing wave boundary condition with the correction up to $\mathcal{O}((r-r_+)^2)$,
\begin{equation}
\label{eq:X_expand}
\begin{split}
    X^\mathrm{in}_{\ell m\omega}(r\rightarrow r_+) = \big[ &1 + a_1(r-r_+) + a_2(r-r_+)^2+ \mathcal{O}((r-r_+)^3)  \big]\mathrm{e}^{-\mathrm{i}kr_\ast}\:,    
\end{split}
\end{equation}
where 
\begin{align}
a_1 &= \frac{(r_+ - M) \left[r_+^2 + (\lambda - 7)M r_+ + 8M^2\right] - 2kMm a^2 r_+^2}{2M r_+ (r_+ - M - 2i k M r_+) (r_+ - M)}\:, \\
a_2 &= \frac{1}{16 M^2 r_+^2 (r_+ - M - i k M r_+) (r_+ - M - 2 i k M r_+) (r_+ - M)^2} \nonumber \\
&\quad \times \Big\{ m^2 a^2 r_+^2 \Big[ (4M^2 k^2 + 2 i k M - 1)r_+^2 - 2 i M (M k + 1) r_+ - M^2 \Big] \nonumber \\
&\quad + 2 m a M r_+ \Big[ -6 k r_+^4 - 2 i r_+^3 (2 - i k M \lambda + 13 i k M) + i M r_+^2 (12 - 2 M^2 k^2 - 2 i k M \lambda + 35 i k M) \nonumber \\
&\quad - 3 i (4 + 5 i k M) M^2 r_+ + 4 i M^3 \Big]  - 2 r_+^6 (1 - 7 i k M) - 2 M r_+^5 (6 - 2 i k M \lambda + 47 i k M - 2 \lambda) \nonumber \\
&\quad + M^2 r_+^4 (156 - 8 i k M \lambda + \lambda^2 + 170 i k M - 26 \lambda)  - 2 M^3 r_+^3 (244 - 2 i k M \lambda - \lambda^2 + 25 i k M - 28 \lambda) \nonumber \\
&\quad + M^4 r_+^2 (678 + \lambda^2 - 104 i k M - 50 \lambda) - 4 M^5 r_+ (111 - 16 i k M - 4 \lambda) + 112 M^6 \Big\}\:.
\end{align}
\end{widetext}
$a_1$ and $a_2$ are derived in Ref.~\cite{Sago:2020avw} by assuming Eq.~\eqref{eq:X_expand} and solving the homogeneous SN equation at each order of $r - r_+$ (see also Refs.~\cite{lo2023recipescomputingradiationkerr} and \cite{nakano2022gravitational} for a review of numerical calculations of the homogeneous solutions). Note that the GW amplitude scales as $\propto \mu/r$ within linear perturbation theory.

\subsection{Post-ISCO gravitational waveforms}
\label{sec:num_GW}
We present the numerical gravitational waveforms here. In the numerical calculation, we obtain $A^\mathrm{inc}_{\ell m\omega}$ and the integral in Eq.~\eqref{eq:X_infty} individually (see Appendix~\ref{app:comp_details} for computational details). Figures~\ref{fig:Post-isco_GWs_q_3} and \ref{fig:Post-isco_GWs_q_5} show the post-ISCO gravitational waveforms for $q=10^{-3}$ (top panels) and $q=10^{-5}$ (bottom panels), respectively. In each row, the blue, red, green, and purple lines represent $h_{22}$, $h_{33}$, $h_{44}$, and $h_{55}$, respectively. Additionally, in each panel, the gray and black lines show the total strain $h_+ = h_{22} + h_{33} + h_{44} + h_{55}$ and $u = u_\mathrm{LR}$.

In Fig.~\ref{fig:Post-isco_GWs_q_3}, unphysical behaviors can be seen before $u = u_\mathrm{LR}$ in the waveforms for $q=10^{-3}$ with $j=0.8$ and $0.9$, especially. These are numerical artifacts resulting from the insufficient simulated time\footnote{Similar phenomena are reported in Ref.~\cite{Lousto:1996sx} (for example, see their Fig.~6). In the inverse Fourier transformation, repetition of the simulated time, denoted by $T_\mathrm{sim}$, is implicitly assumed. If $T_\mathrm{sim}$ is not long enough, the shortage can affect the time-domain counterpart.}. However, it is verified that these do not affect the QNM extraction discussed in Sec.~\ref{sec:evaluation_qnms} by confirming that the fits are performed in a self-consistent manner. These figures indicate that higher harmonic modes are more excited in IMR mergers involving highly spinning BHs compared to EMR mergers. We evaluate the excitations of higher harmonic QNMs in the next section.

\begin{figure*}[t]
\includegraphics[scale=0.47]{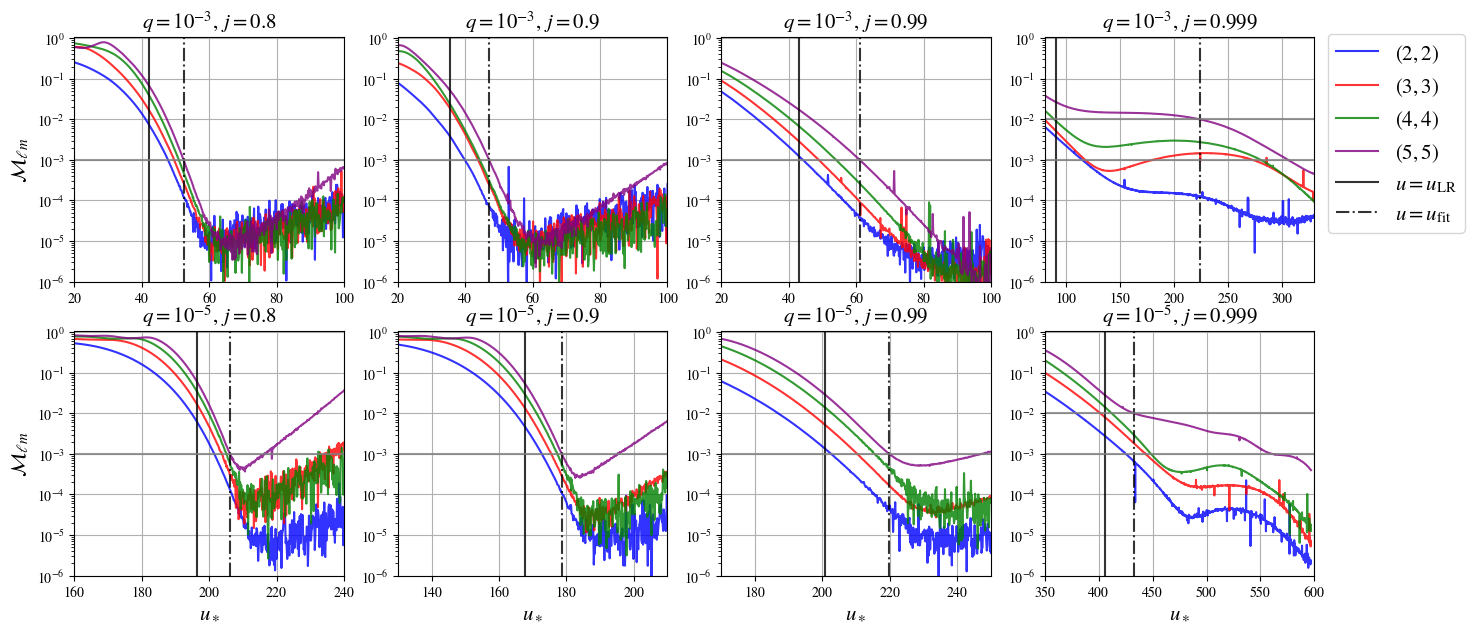}
\caption{\label{fig:Mismatch}
Mismatch $\mathcal{M}_{\ell m}(u_\ast)$ with $q=10^{-3}$ (top) and $q=10^{-5}$ (bottom) for primary BH spins, $j=0.8, 0.9, 0.99,$ and $0.999$ (from left to right).
Black solid and dashed lines show $u=u_\mathrm{LR}$ and $u=u_\mathrm{fit}$, the time at which the QNM model starts working, respectively. $u_\mathrm{fit}$ is defined by the earliest retarded time at which $\mathcal{M}_{\ell m} \leq 0.001$ for $j=0.8, 0.9, 0.99,$ and $\mathcal{M}_{\ell m} \leq 0.01$ for $j=0.999$.}
\end{figure*}

\section{Higher Harmonic QNM Excitation}
\label{sec:higher_harmonic_QNM_excitations}
At sufficiently late times, the waveform is expected to be well described by a superposition of QNMs for a given BH spin. We assess the higher harmonic QNM excitations in the post-ISCO GWs by analyzing their relative amplitudes to the $(\ell, m)=(2, 2)$ mode. To do this, we fit the numerical waveform with a time-domain QNM model that includes up to the first overtone ($n=1$). Several studies (e.g.,~\cite{Baibhav:2023clw}) suggest that extracting more than two overtones from numerical relativity data at a time is challenging, and including many overtones in the model can lead to overfitting. Although it is still unknown if this applies to small-mass-ratio merger waveforms with a highly spinning BH, we follow this approach. Adopting up to $n=1$ allows us to confidently specify the QNM content at late times, owing to fewer fitting parameters, while excluding contributions from higher overtones that can dominate early on~\cite{Oshita:2021iyn, Oshita:2022pkc, Oshita:2022yry, oshita2024reconstructionringdownexcitationfactors}. Instead of identifying contributions from higher overtones, which is challenging, we evaluate the relative amplitude at $u=u_\mathrm{LR}$ as a reference.

In Sec.~\ref{sec:QNM_fit}, we introduce our methodology for fitting the numerically calculated waveform with a time-domain QNM model. Using the QNM model, we evaluate the mismatch, which quantifies the goodness of fit, and determine the time at which the QNM model is considered to start being effective. We closely follow the procedure used in Ref.~\cite{Oshita:2022pkc}. Based on these results, we evaluate the higher harmonic QNM excitations in Sec.~\ref{sec:evaluation_qnms}. Finally, we investigate the inclination angle dependence of the excitations in Sec.~\ref{sec:inclination}.

\begin{figure*}[t]
\centering
\includegraphics[scale=0.45]{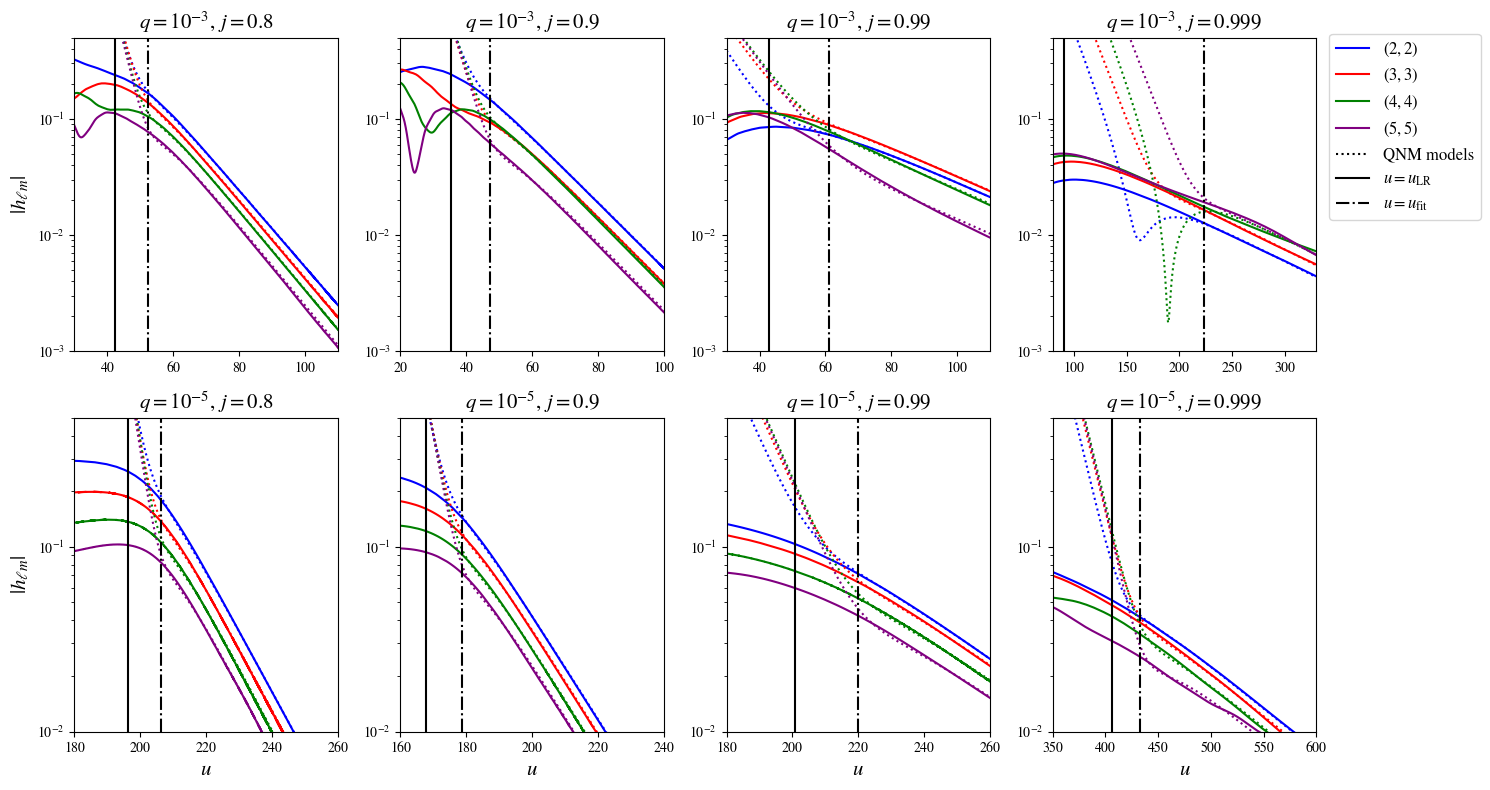}
\caption{\label{fig:amp_num_QNM}
GW amplitudes of the numerical waveforms (solid) and the best-fit QNM models (dotted) for mass ratios of $q=10^{-3}$ (top) and $q=10^{-5}$ (bottom), with primary BH spins of $j=0.8$, $0.9$, $0.99$, and $0.999$ (from left to right). Black solid and dashed lines indicate $u=u_\mathrm{LR}$ and $u=u_\mathrm{fit}$, respectively. The higher harmonic QNM excitations are evaluated at $u=u_\mathrm{fit}$, with $u=u_\mathrm{LR}$ used as a reference (see Fig.~\ref{fig:amp_ratio}).
} 
\end{figure*}

\subsection{QNM model fit in the time domain}
\label{sec:QNM_fit}
\subsubsection{Methodology}
\label{sec:methodology}
We fit a QNM model to the numerical waveform in the time domain for each multipole mode. The QNM model $h^{\mathrm{(QNM)}}_{\ell m}$ is given by
\begin{equation}
\label{eq:qnm_TD}
        h^{\mathrm{(QNM)}}_{\ell m}(u) = \sum_{n=0,1} A_{\ell mn}\mathrm{e}^{-\mathrm{i}\omega_{\ell mn}(u-u_\ast) + \mathrm{i}\phi_{\ell mn}} \theta(u-u_\ast)\:,
\end{equation}
where $u_\ast$ is the retarded time at which the QNM model is assumed to start working, $\omega_{\ell mn}$ is the QNM frequency for the $(\ell, m, n)$ mode, $A_{\ell mn}$ and $\phi_{\ell mn}$ are the real-valued amplitude and phase of the mode at $u=u_\ast$, and $\theta(u)$ is the step function. Throughout this study, we use \texttt{QNM}, a Python module to calculate the QNM frequencies $\omega_{\ell mn}$ for a given BH spin $j$~\cite{Stein:2019mop}. Here, $\{A_{\ell mn},\phi_{\ell mn}, u_\ast\}$ are the fitting parameters for our numerically calculated waveform. They are determined based on the least-squares criteria using \texttt{Fit} available on \textit{Mathematica}. 

We then assess the goodness-of-fit by evaluating the mismatch for each multipole mode, denoted by $\mathcal{M}_{\ell m}$, which is defined by
\begin{equation}
\begin{split}
    &\mathcal{M}_{\ell m}(u_\ast) \\ &:= 1 -\frac{\left|(h_{\ell m}(u)|h^\mathrm{(QNM)}_{\ell m}(u))\right|}{\sqrt{\left(h_{\ell m}(u)|h_{\ell m}(u)\right) (h^\mathrm{(QNM)}_{\ell m}(u)|h^\mathrm{(QNM)}_{\ell m}(u))}}\:,  
\end{split}
\end{equation}
where $[A(u)|B(u)]$ is the inner product between arbitrary functions $A(u)$ and $B(u)$, depending on $u_{\ast}$, and is defined by
\begin{equation}
    \left(A(u)|B(u)\right) := \int^{\infty}_{u_\ast} A^\ast(u) B(u) \diff u\:.
\end{equation}
The superscript ${}^{\ast}$ denotes the complex conjugate. From this procedure, we obtain the best-fit parameter set $\{A_{\ell mn}, \phi_{\ell mn}\}$ and the mismatch $\mathcal{M}_{\ell m}$ for each $u_{\ast}$. In Appendix~\ref{app:behavior_amp}, we confirm that the fits are done in a self-consistent manner by observing the behavior of the best-fit amplitudes.

\subsubsection{Result}
\label{sec:result}
Figure~\ref{fig:Mismatch} shows $\mathcal{M}_{\ell m}(u_\ast)$ for mass ratios $q=10^{-3}$ (top) and $q=10^{-5}$ (bottom) with primary BH spins of $j=0.8$, $0.9$, $0.99$, and $0.999$ (from left to right). Note that the scale of the horizontal axis for $j=0.999$ is different from the others. We define $u_\mathrm{fit}$ as the earliest time at which $\mathcal{M}_{\ell m} \leq 0.001$ for all multipole modes for $j=0.8$, $0.9$, and $0.99$, and $\mathcal{M}_{\ell m} \leq 0.01$ for $j=0.999$. The reason we use $\mathcal{M}_{\ell m} = 0.01$ as the reference for $j=0.999$ is that we have confirmed that only the fundamental mode's amplitude is stably extracted in both cases when using $\mathcal{M}_{\ell m}=0.01$ and $0.001$\footnote{Extracting the first overtone is not possible because $\omega_{\ell m 0}$ and $\omega_{\ell m 1}$ are too close in value.}, and numerical error tends to become more dominant as $u$ increases. In addition, Fig.~\ref{fig:Mismatch} shows a tendency for the rate of decrease of $\mathcal{M}_{\ell m}$ with respect to $u_\ast$ to slow down as the BH spin increases. This can be physically understood by the fact that the decay times of overtones become longer, requiring us to wait until the QNM model, including the first overtone, can accurately describe the waveform.

The amplitude of the numerical waveform and the corresponding best-fit QNM model for each case is illustrated in Fig.~\ref{fig:amp_num_QNM}. The bottom panels show the EMR cases, where the results indicate a clear hierarchy: the $(2, 2)$ mode is the most dominant, and the contribution decreases as $\ell = m$ decreases, similar to comparable mass ratio mergers. On the other hand, as shown in the top panels, this tendency holds only for $j=0.8$ (and subtly for $j=0.9$) in the IMR cases. The higher harmonic modes are more excited than the $(2,2)$ mode in the IMR cases for $j=0.99$ and $0.999$.

\begin{figure*}[t]
\includegraphics[scale=0.47]{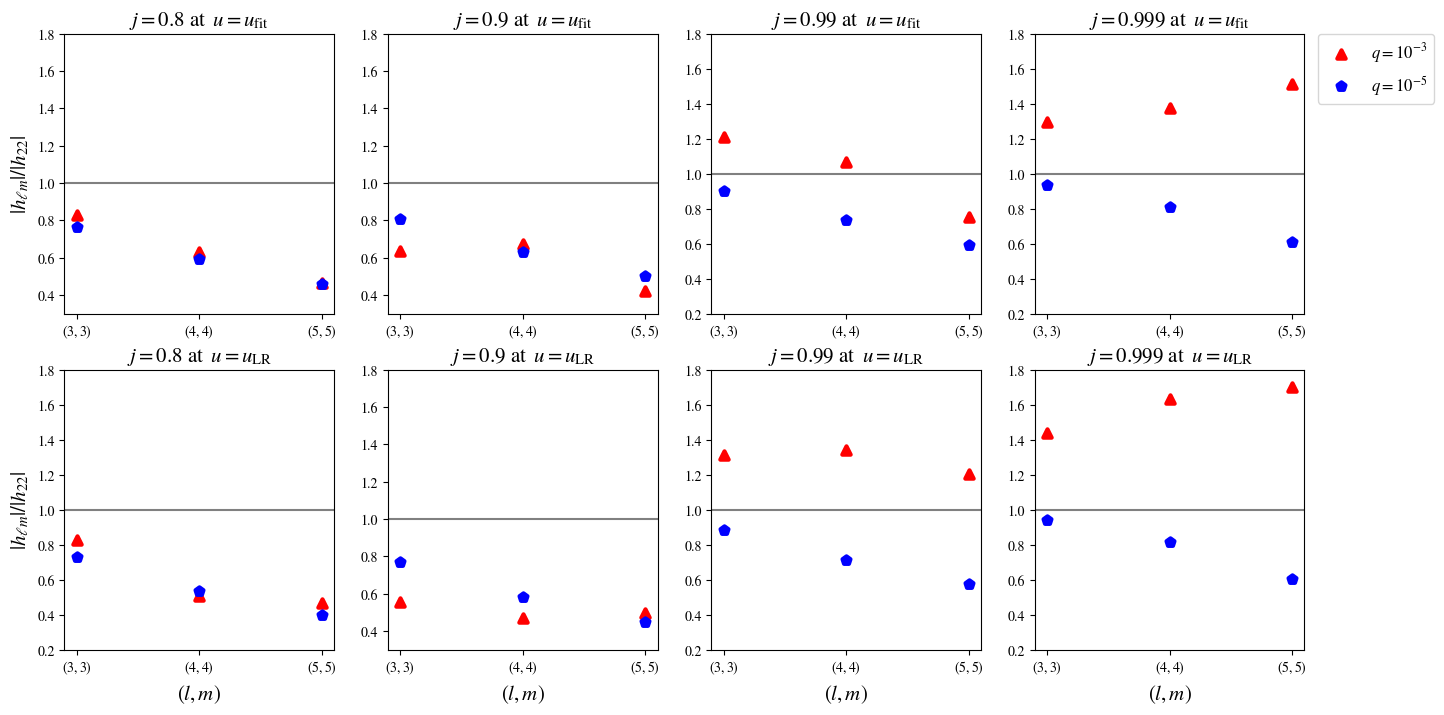}
\caption{\label{fig:amp_ratio}
Relative amplitudes of $(\ell, m)=(3,3), (4,4),$ and $(5,5)$ modes to the $(2,2)$ mode for $j=0.8, 0.9, 0.99,$ and $0.999$ (from left to right) at $u=u_\mathrm{fit}$ (top) and $u = u_\mathrm{LR}$ (bottom). Each panel shows the amplitudes with $q=10^{-3}$ by red marks and $q=10^{-5}$ by blue marks.
}
\end{figure*}

\subsection{Evaluation of higher harmonic QNM excitation}
\label{sec:evaluation_qnms}
We assess the excitations of the higher harmonic QNMs by evaluating the relative amplitudes of the higher harmonic modes compared to the $(2,2)$ mode\footnote{Although there is another indicator of excitation, such as the energy radiation assigned to a mode, we consider the relative amplitude, which is relevant to GW data analysis~\cite{Berti:2005ys}.}, which is the most dominant mode in comparable mass ratio mergers. The results are shown in Fig.~\ref{fig:amp_ratio}. Here, we evaluate the relative amplitude of the higher harmonic modes to the $(2,2)$ mode at $u=u_\mathrm{fit}$ (top panels) and at $u=u_\mathrm{LR}$ (bottom panels) for reference. The red and blue marks correspond to the IMR and EMR cases, respectively.

The $(2,2)$ mode is the most dominant in EMR mergers and in IMR mergers with $j=0.8$ and $j=0.9$. However, as shown in Fig.~\ref{fig:amp_ratio}, even at late times, higher harmonic QNMs are more excited than the $(2,2)$ mode in IMR mergers with $j=0.99$ and $0.999$. Notably, the relative amplitudes of the $(3,3)$ and $(4,4)$ modes exceed unity in IMR mergers with $j=0.99$, and the amplitude increases as $\ell = m$ increases in the case of $j=0.999$. To find the mode that gives the maximum amplitude, we evaluate the GW mode amplitudes with $6 \leq \ell=m \leq 10$ at $u=u_\mathrm{LR}$ in Appendix~\ref{app:additional_calculation}. The result shows that the $(5, 5)$ mode gives the maximum, and the mode amplitude decreases as $\ell$ increases. These results indicate that the mass ratio is crucial in higher harmonic QNM excitations, and the IMR mergers involving near-extremally rotating but still astrophysically relevant BHs efficiently excite higher harmonic QNMs.

A comparison of the top and bottom panels for $j=0.99$ and $0.999$ in Fig.~\ref{fig:amp_ratio} reveals that the amplitudes evaluated at $u=u_\mathrm{LR}$ are larger than those at $u=u_\mathrm{fit}$. Since the damping time of each $(\ell, m)$ mode with the same $n$ converges to the same value as $j\rightarrow 1$, this might indicate that higher overtones are excited early in the higher multipole modes. Although it is practically difficult to identify these modes using the fitting procedure, a possible way to resolve these overtones will be discussed in Sec.~\ref{sec:discussion}.

Additionally, we have confirmed that varying both $\Delta \hat{L}$ and $\Delta \hat{E}$ by $1\:\%$ from Eqs.~\eqref{eq:Delta_L} and \eqref{eq:Delta_E} affects these relative amplitudes by only $< 3.6\:\%$ and, importantly, does not change the hierarchy of the relative amplitudes. However, this solely indicates the insensitivity of the excitations to small modifications in $\Delta \hat{L}$ and $\Delta \hat{E}$, and we cannot conclude the validity of the Ori-Thorne procedure itself from this result. Several discussions on this point will be presented in Sec.~\ref{sec:discussion}.

\begin{figure*}[t]
\includegraphics[scale=0.47]{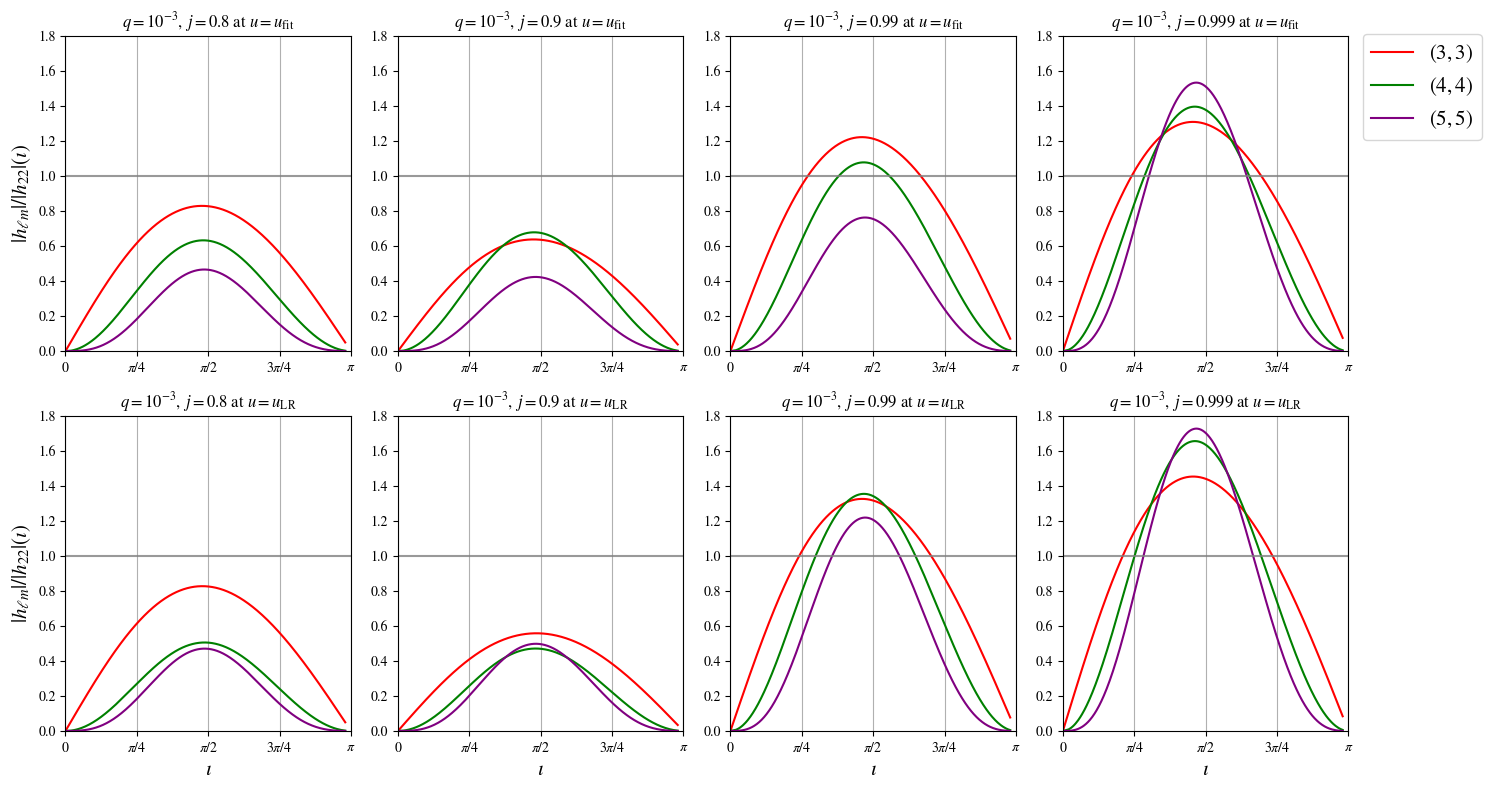}
\caption{\label{fig:inclination_dependence}
Inclination angle dependence of the relative amplitudes to the $(2,2)$ mode of the $(\ell, m)=(3,3)$, $(4,4)$, and $(5,5)$ modes in IMR mergers for $j=0.8$, $0.9$, $0.99$, and $0.999$ (from left to right) at $u=u_\mathrm{fit}$ (top) and $u = u_\mathrm{LR}$ (bottom). The values at $\iota=\pi/2$ correspond to those in Fig.~\ref{fig:amp_ratio}.
}
\end{figure*}

\subsection{Dependence on inclination angle}
\label{sec:inclination}
In Sec.~\ref{sec:evaluation_qnms}, we evaluated the excitations of higher harmonic QNMs assuming an inclination angle of $\iota=\pi/2$, i.e., an edge-on binary. However, the relative amplitudes among harmonic modes depend on the inclination angle $\iota$ through the spin-weighted spheroidal harmonic ${}_{-2}S^{a\omega}_{\ell m}(\iota)$.

Figure~\ref{fig:inclination_dependence} shows the $\iota$-dependence of the relative amplitudes of the $(\ell, m)=(3,3)$, $(4,4)$, and $(5,5)$ modes relative to the $(2,2)$ mode in IMR mergers for $j=0.8$, $0.9$, $0.99$, and $0.999$ (from left to right) at $u=u_\mathrm{fit}$ (top) and $u = u_\mathrm{LR}$ (bottom). In each panel, the red, green, and purple lines depict the relative amplitudes of the $(3,3)$, $(4,4)$, and $(5,5)$ modes, respectively. The values at $\iota=\pi/2$ correspond to those in Fig.~\ref{fig:amp_ratio}. These results suggest that binaries with $\pi/4 \leq \iota \leq 3\pi/4$ (around edge-on) are preferred since the higher harmonic QNMs tend to be more informative.

Notably, the results are consistent with one of the analytical calculations of Ref.~\cite{Compere:2017hsi}, which calculates GWs from several types of orbits (including plunges) around nearly maximally spinning BHs in the near-horizon extremal Kerr (NHEK) region. In Ref.~\cite{Compere:2017hsi}, it is shown that modes up to $\ell=20, m=20$ should be included to have a $5\:\%$ accuracy of the energy emission for the edge-on observer radiated by the plunge, which involves the nearly maximally spinning BH and the secondary object with the energy and angular momentum whose values are those at the ISCO. While clarifying the correspondence between the NHEK and our treatment is necessary, this also supports the result that the edge-on binaries involving rapidly spinning primary BHs can efficiently radiate higher harmonic QNMs.

\section{Discussion}
\label{sec:discussion}
What do the results indicate for future GW observations by space-based interferometers? A key to addressing this question is an asymptotic behavior observed in most QNM frequencies for the $(\ell, m)$ mode with $\ell = m$: $2M\omega_{\ell m n} \rightarrow m$ as $j \rightarrow 1$. This means that as the BH spin increases, distinguishing between overtones with the same $(\ell,m)$ becomes more challenging, making it more necessary to identify higher harmonic QNMs in the signal. Moreover, due to the higher real frequency of a higher harmonic mode, more precise parameter estimation is expected if it is sufficiently excited and informative. Therefore, provided that both observation and theory suggest that BHs with masses of $\sim 10^7 M_\odot$ are rapidly spinning, these excitations are beneficial for the precise inference of BH properties or for the BH spectroscopy program using space-based GW interferometers.

Another interesting idea is to compare the results of the ringdown-only analysis with those of the inspiral-only analysis. In the context of observational tests of GR with GWs, these excitations will play a crucial role, especially in the inspiral-merger-ringdown consistency test, which checks the consistency of the estimated results between the inspiral-only and post-inspiral-only analyses (e.g.,~\cite{LIGOScientific:2016lio, Ghosh:2016qgn, LIGOScientific:2019fpa, Ghosh:2017gfp, LIGOScientific:2021sio_a}). 
In general, ringdown GWs can be more informative in IMR mergers than in EMR mergers since the leading-order GW amplitude in mass ratio scales as $\propto \mu/r$, and therefore, the analysis of IMR merger data is likely to provide tighter constraints. Importantly, in addition to that expectation, the higher harmonic QNM excitations will further enhance the measurement precision in the inspiral-merger-ringdown consistency test.

In our calculation, while the mass ratio effect is incorporated through the radiation fluxes during the transition, the GW calculations are performed for geodesic motions. Therefore, one direct interpretation is that the higher harmonic QNM excitation is sensitive to the secondary object’s energy, angular momentum, and the primary BH’s spin. Since it is known that the $(\ell, m) = (2,2)$ mode is dominant in the comparable mass BH mergers as well as in the EMR mergers shown in this study, our results suggest the existence of an optimal energy and angular momentum configuration for the higher harmonic QNM excitation. 

In addition, several studies derive analytic expressions of the post-ISCO ringdown GW amplitudes for near extremally spinning BHs using the NHEK approximation, which enhances the spacetime symmetry~(e.g.~\cite{Hadar:2014dpa, Hadar:2015xpa, Hadar:2016vmk, Compere:2017hsi}). In our calculation, we do not use such an approximation and derive the GWs numerically for the BH spin cases including the moderate spins within the same numerical calculation framework. A comparison between the analytic evaluation using the NHEK approximation and our full numerical calculation is left for future work.

We end by providing some comments on the main assumptions made in this study.
\begin{itemize}
    \item We follow the Ori-Thorne procedure to estimate the energy and angular momentum losses in the transition regime as a simple approach. As long as we adopt this procedure, we have confirmed that the excitations are insensitive to small changes in the radiated energy and angular momentum in the transition regime. However, this approach becomes subtle for near-extremally rotating BHs since there is another small parameter, $\lambda:=\sqrt{1-j^2}$, which contributes to the equation of motion in addition to $q$~\cite{Compere:2019cqe}. The Ori-Thorne procedure is valid when $q \ll \lambda$. For $j=0.999$, corresponding to $\lambda = 0.0447$, we have $\lambda/q = 44.7$, which is larger than unity but not by a significant amount (whereas for $j=0.99$, $\lambda/q = 141$). Furthermore, Ref.~\cite{Kesden:2011ma} points out that Eqs.~\eqref{eq:Delta_L} and \eqref{eq:Delta_E}, which approximate energy and angular momentum radiation in the transition regime, may overestimate these values for a highly spinning BH. Therefore, more accurate considerations could reduce the excitations. On the other hand, the hierarchy is expected to hold, as the results for $j=0.8, 0.9,$ and $0.99$ indicate that higher harmonic QNMs are more excited in IMR mergers as $j$ increases. Investigating how more appropriate treatments affect the excitations is left for future study.

    \item We adopt the QNM model including up to $n=1$ for the fit to confidently specify the components of the long-lived modes, which means that we cannot expect reliable extractions of the higher overtones. However, Fig.~\ref{fig:amp_ratio} suggests that the higher harmonics of higher multipole modes may be excited at early times. While it is practically difficult to extract more than two overtones through simultaneous fitting~\cite{Baibhav:2023clw}, Ref.~\cite{takahashi2023iterativeextractionovertonesblack} recently proposed an iterative extraction method, demonstrating that it allows for the extraction of multiple overtones from numerical relativity data. Applying this method to our numerical waveforms is an interesting direction for future work.

\end{itemize}

\section{Conclusion}
\label{sec:summary_and_conclusion}
Given that higher harmonic QNM excitations play crucial roles in the precise measurement of BH properties and BH spectroscopy with LISA, we numerically evaluate the higher harmonic QNM excitations in the post-ISCO GWs of rapidly spinning BHs and investigate the mass ratio dependence of these excitations. We consider BH spins of $j=0.8, 0.9, 0.99,$ and $0.999$ and mass ratios of $q=10^{-3}$ (IMR) and $10^{-5}$ (EMR). The effect of the mass ratio is accounted for by considering the gravitational radiation during the transition regime, i.e., before the plunge, following the Ori-Thorne procedure. The waveforms are numerically calculated based on the Teukolsky-Sasaki-Nakamura formalism. To assess the excitations of higher harmonic QNMs, we first fit the waveform with a time-domain QNM model, including up to the first overtone for each multipole mode, and evaluate $\mathcal{M}_{\ell m}$, which quantifies the goodness-of-fit. Based on the value of $\mathcal{M}_{\ell m}$, we determine the time at which the QNM model is valid and derive the relative amplitude of each multipole mode compared to the $(2,2)$ mode at that time as an indicator of the excitation.

Our main result is that higher harmonic QNMs are more excited in IMR mergers involving a highly spinning but still astrophysically relevant BH than the $(2,2)$ mode, which is the dominant mode for comparable mass ratio mergers. This is advantageous for the inference of SMBH properties and the BH spectroscopy program, given that the confident detection of multiple overtones becomes more difficult as the BH spin increases due to the closeness of the QNM frequencies. In addition, such detection will provide a privileged opportunity for the inspiral-merger-ringdown consistency test owing to these excitations.

As discussed in Sec.~\ref{sec:discussion}, we adopt the Ori-Thorne procedure as a simple method to approximate the energy and angular momentum radiation in the transition regime; however, more general treatments can be found in the literature (e.g.,~\cite{Kesden:2011ma, Apte:2019txp}). The difference becomes significant for near-extremal BHs~\cite{Compere:2019cqe}. In our study, given that the Ori-Thorne procedure is valid if $q \ll \lambda$, where $\lambda := \sqrt{1-j^2}$, the case with $q=10^{-3}$ for $j=0.999$ is subtle. Considering these methods may reduce the excitations, although the trend that higher harmonic QNMs become more excited in IMR mergers as the BH spin increases is expected to hold.

\begin{acknowledgments}
I gratefully acknowledge Naritaka Oshita and Norichika Sago for helpful comments and discussions, and Hiroyuki Nakano for suggesting relevant references. 
I thank Kipp Cannon, Yanbei Chen, Kenta Hotokezaka, Soichiro Morisaki, Hayato Motohashi, Hiroki Takeda, Daichi Tsuna, and Atsushi Nishizawa for fruitful discussions, and Geoffrey Compère for valuable comments on the earlier version of the manuscript. This work makes use of the Black Hole Perturbation Toolkit. I am supported by JSPS KAKENHI grant No. 23KJ06945.
\end{acknowledgments}

\appendix
\begin{widetext}
\section{SOURCE TERM OF THE SASAKI-NAKAMURA EQUATION FOR A GENERIC GEODESIC ORBIT}
\label{app:SN_source}
We show the explicit form of the source term $\mathcal{S}_{\ell m\omega}$ for a generic geodesic motion. Here, we denote ${}_{-2}S^{a\omega}_{\ell m}$ by $S$ for brevity. While Ref.~\cite{Saijo:1996iz} presents the form in Appendix~A, our expression is partially different due to how to perform partial integration\footnote{Apart from the difference, Ref.~\cite{Saijo:1996iz} seems to contain typos.}. We derive the explicit expression such that our result reduces to the one presented in Refs.~\cite{Kojima:1984cj, Nakamura:1987zz} when taking $\hat{E} = 1$.

We start with the equation of motion for a generic geodesic orbit,
\begin{align}
    \label{eq:t_gen}
    &\Sigma\frac{\diff t}{\diff \tau} = -a(a\hat{E}\sin^2\theta-\hat{L}) + \frac{r^2+a^2}{\Delta}P\:,\\
    \label{eq:r_gen}
    &\Sigma \frac{\diff r}{\diff \tau} = \pm\sqrt{R}\:,\\
    \label{eq:theta_gen}
    &\Sigma \frac{\diff \theta}{\diff \tau} = \pm \sqrt{\Theta}\:,\\
    \label{eq:phi_gen}
    &\Sigma \frac{\diff \phi}{\diff \tau} = -\left( a\hat{E}-\frac{\hat{L}}{\sin^2\theta} \right) + \frac{a}{\Delta}P\:,
\end{align}
where
\begin{align}
    P &= \hat{E} (r^2+a^2) - a\hat{L}\:,\\
    R &= P^2 - \Delta(r^2 + (\hat{L}-a\hat{E})^2 + C)\:,\\
    \Theta &= C - \cos^2\theta \left[ a^2(1-\hat{E}^2) + \frac{\hat{L}^2}{\sin^2\theta} \right]\:,
\end{align}
with $C$ is the Carter constant.
Note that in this study, we consider the equatorial motion, $\theta(t)=\pi/2$, which enforces $C=0$.

$B'_2$ and $B'^*_2$ in Eq.~\eqref{eq:T_lmw} are given by
\begin{align}
B'_2 &= -\frac{1}{2} \rho^8 \Bar{\rho} \hat{L}_{-1} \left[ \rho^{-4} \hat{L}_0 \left( \rho^{-2} \Bar{\rho}^{-1} T_{nn} \right) \right] -\frac{1}{2\sqrt{2}} \rho^8 \Bar{\rho} \Delta^2 \hat{L}_{-1} \left[ \rho^{-4} \Bar{\rho}^2 \hat{J}_+ \left( \rho^{-2} \Bar{\rho}^{-2} \Delta^{-1} T_{mn} \right) \right]\:,\\
B'^*_2 &= -\frac{1}{4} \rho^8 \Delta^2 \hat{J}_+ \left[ \rho^{-4} \Bar{\rho}^2 \hat{J}_+ \left( \rho^{-2} \Bar{\rho} T_{mm} \right) \right] -\frac{1}{2\sqrt{2}} \rho^8 \Delta^2 \hat{J}_+ \left[ \rho^{-4} \Bar{\rho}^2 \Delta^{-1} \hat{L}_{-1} \left( \rho^{-2} \Bar{\rho}^{-2} T_{mn} \right) \right]\:,
\end{align}
where 
\begin{align}
    \hat{L}_s  &= \frac{\partial}{\partial \theta} - \frac{\mathrm{i}}{\sin\theta} \frac{\partial}{\partial \phi} - \mathrm{i}\sin\theta\frac{\partial}{\partial t} + s\cot\theta\:,\\
    \hat{J}_+ &= \frac{\partial}{\partial r} - \left( \frac{r^2+a^2}{\Delta}\frac{\partial}{\partial t} + \frac{a}{\Delta}\frac{\partial}{\partial \phi} \right)\:,
\end{align}
and $T_{nn}, T_{\overline{m}n},$ and $T_{\overline{mm}}$ are the tetrad components of the energy-momentum tensor, which is given by
\begin{equation}
\begin{split}
    T^{\alpha\beta} = &\frac{\mu}{\Sigma\sin\theta \frac{\diff t}{\diff \tau}} \frac{\diff z^\alpha}{\diff \tau} \frac{\diff z^\beta}{\diff \tau} \delta \left( r-r(t) \right) \delta \left( \theta-\theta(t) \right) \delta \left( \phi-\phi(t) \right)\:,
\end{split}
\end{equation}
where $z^\alpha=(t, r(t), \theta(t), \phi(t))$ is a geodesic trajectory governed by Eqs.~\eqref{eq:t_gen}--\eqref{eq:phi_gen}. Explicit forms of $T_{nn}, T_{\overline{m}n},$ and $T_{\overline{mm}}$ are given by
\begin{align}
    &T_{nn} = \mu \frac{C_{nn}}{\sin\theta} \delta \left( r-r(t) \right) \delta \left( \theta-\theta(t) \right)  \delta \left( \phi-\phi(t) \right)\:,\\
    &T_{\overline{m}n} = \mu \frac{C_{\overline{m}n}}{\sin\theta} \delta \left( r-r(t) \right)  \delta \left( \theta-\theta(t) \right) \delta \left( \phi-\phi(t) \right) \:,\\
    &T_{\overline{mm}} = \mu \frac{C_{\overline{mm}}}{\sin\theta} \delta \left( r-r(t) \right)  \delta \left( \theta-\theta(t) \right) \delta \left( \phi-\phi(t) \right) \:,
\end{align}
where 
\begin{align}
    C_{nn} &= \frac{1}{4\Sigma^3} \left( \frac{\diff t}{\diff \tau} \right)^{-1} \left( P + \Sigma \frac{\diff r}{\diff \tau} \right)^2\:,\\
    C_{\overline{m}n} &= -\frac{\rho}{2\sqrt{2} \Sigma^2} \left( \frac{\diff t}{\diff \tau} \right)^{-1} \left( P + \Sigma \frac{\diff r}{\diff \tau} \right)\left[\mathrm{i}\sin\theta \left(a\hat{E}-\frac{\hat{L}}{\sin\theta}\right)+\Sigma\frac{\diff \theta}{\diff \tau}\right] \:,\\
    C_{\overline{mm}} &= -\frac{\rho^2}{2\Sigma} \left( \frac{\diff t}{\diff \tau} \right)^{-1} \left[ \mathrm{i}\sin\theta\left(a\hat{E}-\frac{\hat{L}}{\sin\theta}\right) + \Sigma\frac{\diff \theta}{\diff \tau}\right]^2\:.
\end{align}
Using an identity for arbitrary functions $a(\theta)$ and $b(\theta)$
\begin{equation}
    \int^{\pi}_0 a(\theta) L_s[b(\theta)] \sin\theta \diff \theta = -\int^\pi_0 b(\theta) L^\dagger_{1-s}[a(\theta)] \sin\theta \diff \theta\:,
\end{equation}
to perform the partial integral, Eq.~\eqref{eq:T_lmw} can be expressed as
\begin{equation}
\label{eq:T_lmw_2}
    \begin{split}
        T_{\ell m\omega} =\:\frac{4\mu}{\sqrt{2\pi}} &\int^\infty_{-\infty} \diff t \int \diff \theta \:\mathrm{e}^{\mathrm{i}\omega t -\mathrm{i}m\phi(t)}\\
        &\:\times\Bigg\{ -\frac{1}{2}L^\dagger_1\left[ \rho^{-4} L^\dagger_2 (\rho^3 S) \right] C_{nn}\rho^{-2}\Bar{\rho}^{-1} \delta(r-r(t))\delta(\theta-\theta(t)) \\
        &\:\:\:\:\:\:\:\:\:\:+\frac{\Delta^2 \bar{\rho}}{\sqrt{2}\rho^2} L^\dagger_2 (\rho\Bar{\rho}S) J_+\left[ \frac{C_{\overline{m}n}}{\rho^2\Bar{\rho}^2\Delta}\delta(r-r(t))\delta(\theta-\theta(t)) \right] \\
        &\:\:\:\:\:\:\:\:\:\:+\frac{1}{2\sqrt{2}}L^\dagger_{2}\left[ \rho^3 S\frac{\partial }{\partial r}(\Bar{\rho}^2 \rho^{-4})\right] \frac{C_{\overline{m}n}\Delta}{\rho^{2}\Bar{\rho}^{2}} \delta(r-r(t))\delta(\theta-\theta(t))\\
        &\:\:\:\:\:\:\:\:\:\:-\frac{1}{4}\rho^3 \Delta^2 S J_+\left[ \rho^{-4}J_+(\Bar{\rho}\rho^{-2} C_{\overline{mm}}\delta(r-r(t))\delta(\theta-\theta(t)) ) \right]
        \Bigg\}\:,
    \end{split}
\end{equation}
where
\begin{align}
    &L^\dagger_s = \frac{\partial}{\partial \theta} - \frac{m}{\sin\theta} + a\omega \sin\theta + s\cot\theta\:,\\
    & J_+ = \frac{\partial}{\partial r} + \mathrm{i}\frac{K}{\Delta}\:.    
\end{align}
$\mathcal{W}$ in Eq.~\eqref{eq:S_lnw} is related to $T_{\ell m\omega}$ by
\begin{equation}
\label{eq:diff_W}
    \frac{\diff^2 \mathcal{W} }{\diff r^2} = -\frac{r^2}{\Delta^2}T_{\ell m\omega}\:\mathrm{exp}\left[ \mathrm{i}\int^r \frac{K}{\Delta} \diff r \right]\:.
\end{equation}
Dividing $\mathcal{W}$ in Eq.~\eqref{eq:diff_W} into three parts as
\begin{equation}
    \mathcal{W} = \mathcal{W}_{nn} + \mathcal{W}_{\overline{mm}} + \mathcal{W}_{\overline{m}n}\:,
\end{equation}
we can find that $\mathcal{W}_{nn}, \mathcal{W}_{\overline{mm}},$ and $\mathcal{W}_{\overline{m}n}$ satisfy the following equations:
\begin{equation}
    \frac{\sqrt{2\pi}}{\mu}\frac{\diff^2 \mathcal{W}_{nn}}{\diff r^2} = \frac{r^2}{2\rho\Delta^2}\left| \frac{\diff r}{\diff \tau} \right| \left( 1 - \frac{P}{\sqrt{R}} \right)^2 L^\dagger_1\left[ \rho^{-4} L^\dagger_2 (\rho^3 S) \right] \mathrm{e}^{\mathrm{i}\chi}\:, \label{eq:diff_W_nn}
\end{equation}

\begin{equation}
\begin{split}
        \frac{\sqrt{2\pi}}{\mu}\frac{\diff^2 \mathcal{W}_{\overline{m}n}}{\diff r^2} &= \int^{\infty}_{-\infty} \diff t \:\mathrm{e}^{\mathrm{i}\omega t-\mathrm{i}m\phi(t)}\left\{ -\frac{r^2 \Bar{\rho}}{\rho^2} L^\dagger_2(\rho\Bar{\rho}S)J_+\left[ \rho \frac{\diff r}{\diff t} \frac{\Sigma}{\Delta} \left( 1 - \frac{P}{\sqrt{R}}  \right)w^{(1)}_{\overline{m}n} \delta(r-r(t)) \right]\mathrm{e}^{\mathrm{i}\int^r \frac{K}{\Delta} \diff r} \right\}\\
        &\:\:\:\:\:\:- \mathrm{sgn}\left( \frac{\diff r}{\diff \tau} \right) \frac{r^2 \rho}{2}L^\dagger_2 \left[ \rho^3 S (\Bar{\rho}^2\rho^{-4})' \right] \frac{\Sigma}{\Delta} \left( 1-\frac{P}{\sqrt{R}} \right) w^{(1)}_{\overline{m}n} \mathrm{e}^{\mathrm{i}\chi}\:,\label{eq:diff_W_mn}
\end{split}
\end{equation}
\begin{equation}
\label{eq:diff_W_mm}
    \frac{\sqrt{2\pi}}{\mu} \frac{\diff^2 \mathcal{W}_{\overline{mm}}}{\diff r^2} = \int^\infty_{-\infty} \diff t \:\mathrm{e}^{\mathrm{i}\omega t-\mathrm{i}m\phi(t)} S 
    \left\{ r^2\rho^3 J_+ 
    \left[ \rho^{-4} J_+ 
    \left( \frac{\rho\Bar{\rho}^2}{2} \left( \frac{\diff t}{\diff \tau} \right)^{-1} \delta(r-r(t)) \left(w^{(1)}_{\overline{m}n}\right)^2 \right) \right] 
    \mathrm{e}^{\mathrm{i}\int^r \frac{K}{\Delta} \diff r}\right\}\:,
\end{equation}
where $\chi$ and $w_{\overline{m}n}^{(1)}$ is defined by
\begin{align}
    \chi &= \omega \tilde{V} - m\tilde{\phi}\:, \\
    w^{(1)}_{\overline{m}n} &= -\left[ \pm\sqrt{\Theta} + \mathrm{i}\sin\theta\left( a\hat{E}-\frac{\hat{L}}{\sin\theta} \right) \right]\:,
\end{align}
with 
\begin{align}
    \tilde{V} &= t + r_\ast\:,\\
    \tilde{\phi} &= \phi + \int^{r} \frac{a}{\Delta} \diff r\:.
\end{align}
We can solve Eqs.~\eqref{eq:diff_W_nn} to \eqref{eq:diff_W_mm} integrating these equations by parts twice. 
The solutions are given in the following forms:
\begin{align}
    &\frac{\sqrt{2\pi}}{\mu}\mathcal{W}_{nn}(r)=\:f_0(r)\:\mathrm{e}^{\mathrm{i}\chi(r)} + \int^\infty_{r} f_1(r_1)\:\mathrm{e}^{\mathrm{i}\chi(r_1)}\diff r_1 + \int^\infty_{r} \diff r_1 \int^\infty_{r_1} f_2(r_2)\:\mathrm{e}^{\mathrm{i}\chi(r_2)}\diff r_2\:,\\
    &\frac{\sqrt{2\pi}}{\mu}\mathcal{W}_{\overline{m}n}(r) =\:g_0(r)\:\mathrm{e}^{\mathrm{i}\chi(r)} + \int^\infty_{r} g_1(r_1)\:\mathrm{e}^{\mathrm{i}\chi(r_1)}\diff r_1 + \int^\infty_{r} \diff r_1 \int^\infty_{r_1} g_2(r_2)\:\mathrm{e}^{\mathrm{i}\chi(r_2)}\diff r_2\:,\\
    &\frac{\sqrt{2\pi}}{\mu}\mathcal{W}_{\overline{mm}}(r) =\:h_0(r)\:\mathrm{e}^{\mathrm{i}\chi(r)} + \int^\infty_{r} h_1(r_1)\:\mathrm{e}^{\mathrm{i}\chi(r_1)}\diff r_1 + \int^\infty_{r} \diff r_1 \int^\infty_{r_1} h_2(r_2)\:\mathrm{e}^{\mathrm{i}\chi(r_2)}\diff r_2\: \label{eq:W_mm}.
\end{align}
where 
\begin{align}
    \label{eq:f_0}
    f_0 &= -\frac{1}{\omega^2}w_{nn}\:,\\
    f_1 &= -\frac{1}{\omega^2} \left[ w'_{nn} + \mathrm{i}\eta w_{nn}  \right]\:,\\
    f_2 &= \frac{\mathrm{i}}{\omega} \left\{ \left[ w'_{nn} + \mathrm{i}\eta w_{nn} \right] \left( \tilde{V}' - \frac{a(a\hat{E}\sin^2\theta - \hat{L})}{\sqrt{R}} \right) + w_{nn} \left( \tilde{V}' - \frac{a(a\hat{E}\sin^2\theta - \hat{L})}{\sqrt{R}} \right)'  \right\}\:,\\
    g_0 &= \frac{\mathrm{i}}{\omega} w^{(1)}_{\overline{m}n} \frac{w^{(2)}_{\overline{m}n}}{\Bar{\rho}(r^2+a^2)} \mathrm{sgn}\left( \frac{\diff r}{\diff \tau} \right)\:,\\
    g_1 &= \frac{\mathrm{i}}{\omega} w^{(1)}_{\overline{m}n} \left[ -w^{(3)}_{\overline{m}n} + \left( \frac{w^{(2)}_{\overline{m}n}}{\Bar{\rho}(r^2+a^2)} \right)' + \frac{w^{(2)'}_{\overline{m}n}}{\Bar{\rho}(r^2+a^2)} + \mathrm{i}\eta \frac{w^{(2)}_{\overline{m}n}}{\Bar{\rho}(r^2+a^2)} \right] \mathrm{sgn}\left( \frac{\diff r}{\diff \tau} \right)\:,\\
    g_2 &= -\frac{\mathrm{i}}{\omega} w^{(1)}_{\overline{m}n} \left[ \left( w^{(3)}_{\overline{m}n} - \frac{w^{(2)'}_{\overline{m}n}}{\Bar{\rho}(r^2+a^2)} \right)' + \mathrm{i}\eta \left( w^{(3)}_{\overline{m}n} + \frac{w^{(2)'}_{\overline{m}n}}{\Bar{\rho}(r^2+a^2)} \right)  \right] \mathrm{sgn}\left( \frac{\diff r}{\diff \tau} \right)\:,\\
    h_0 &= \frac{r^2\Bar{\rho}^2}{2}\left| \frac{\diff r}{\diff \tau} \right|^{-1} S \left(w^{(1)}_{\overline{m}n}\right)^2\:,\\
    h_1 &= \frac{\rho\Bar{\rho}^2}{2} \left| \frac{\diff r}{\diff \tau} \right|^{-1} S \left(w^{(1)}_{\overline{m}n}\right)^2 \left[ \left(\frac{r^2}{\rho}\right)' + \rho^{-4} \left( \rho^3 r^2 \right)'
 \right]\:,\\
    h_2 &= \frac{\rho\Bar{\rho}^2}{2} \left| \frac{\diff r}{\diff \tau} \right|^{-1} S \left(w^{(1)}_{\overline{m}n}\right)^2 \left[ \rho^{-4}\left( \rho^3 r^2 \right)' \right]'\:,
\end{align}
with
\begin{align}
    \label{eq_w_nn}
    w_{nn} &= \frac{r^2}{2 \rho (r^2+a^2)^2} \left| \frac{\diff r}{\diff \tau} \right| L_1^{\dagger} \left[ \rho^{-4} L_2^{\dagger} (\rho^3 S) \right]\:,\\
    w^{(2)}_{\overline{m}n} &= \frac{r^2 \Bar{\rho}}{\rho^2} L^\dagger_2 \left[ \rho\Bar{\rho} S \right]\:,\\
    w^{(3)}_{\overline{m}n} &= \frac{r^2}{2\rho (r^2+a^2)} L^\dagger_2 \left[ \rho^3 S \left( \Bar{\rho}^2 \rho^{-4} \right)' \right]\:.
\end{align}
Here, $\mathrm{sgn}(\diff r/\diff \tau)$ is a signature of $\diff r/\diff \tau$ and becomes a minus for the case of an infalling particle.
To derive the concrete form of $f_i, g_i$, and $h_i$ $(i=0,1,2)$, we utilize the following identities for an arbitrary function $f(r)$:
\begin{equation}
    \label{eq:identity_1}
    f(r) \left( \tilde{V}' - \frac{a(a\hat{E}\sin^2\theta - \hat{L})}{\sqrt{R}} \right) \mathrm{e}^{\mathrm{i}\chi} = -\frac{\mathrm{i}}{\omega} \left\{ \left[ f(r)\mathrm{e}^{\mathrm{i}\chi} \right]' - \left[f'(r) + \mathrm{i}\eta f(r) \right] \mathrm{e}^{\mathrm{i}\chi} \right\}\:,    
\end{equation}
where
\begin{equation}
\begin{split}
    \eta(r) &= \frac{a\omega (a\hat{E}\sin^2\theta - \hat{L})}{\sqrt{R}} - m\tilde{\phi}'\\ &= \left( a\omega - \frac{m}{\sin^2\theta} \right)\frac{a\hat{E}\sin^2{\theta}-\hat{L}}{\sqrt{R}} - \frac{am}{\Delta} \left( 1 - \frac{P}{\sqrt{R}} \right)\:.
\end{split}
\end{equation}
and
\begin{equation}
    \label{eq:identity_2}
    \left[J_+ f(r)\right] \mathrm{exp}\left[\mathrm{i}\int^r \frac{K}{\Delta}\diff r \right]  =  \left\{ f(r)\mathrm{exp}\left[\mathrm{i}\int^r \frac{K}{\Delta}\diff r \right]\right\}'\:.
\end{equation}
Especially, Eqs~\eqref{eq:identity_1} and \eqref{eq:identity_2} are used in performing partial integration for the $n$ and $m$ components, respectively.

For the equatorial orbits, the terms involving $S$ become
\begin{align}
    &L_1^{\dagger} \left[ \rho^{-4} L_2^{\dagger} (\rho^3 S) \right]\Big|_{\theta=\pi/2} =\:2r \left\{\left(-m+a\omega-\frac{\mathrm{i}a}{r}\right)\left[ S_1 + (-m+a\omega)S_0 \right] - \frac{\lambda}{2}S_0\right\} =: 2r\hat{S}\:,\\
    &L_2^{\dagger} \left[ \rho \Bar{\rho} S \right]\Big|_{\theta=\pi/2} = \frac{1}{r^2} \left[S_1 + (-m + a \omega) S_0\right]\:,\\ 
    &L^\dagger_2 \left[ \rho^3 S \left( \Bar{\rho}^2 \rho^{-4} \right)' \right]\Big|_{\theta=\pi/2} = \frac{2}{r^2} \left[S_1 + (-m + a \omega) S_0\right]\:,
\end{align}
where $S_0 = S(\theta=\pi/2)$, $S_1 = \diff S/\diff \theta (\theta=\pi/2)$, and $\hat{S}$ is the same as the one shown in Appendix B of Ref.~\cite{Kojima:1984cj}. 
\end{widetext}

\section{COMPUTATIONAL DETAILS}
\label{app:comp_details}
To numerically derive $A^\mathrm{inc}_{\ell m\omega}$ and the integral in Eq.~\eqref{eq:X_infty}, we use an equally spaced grid with $N$ divisions for ${r_\ast}_\mathrm{min} \leq r_\ast \leq {r_\ast}_\mathrm{max}$. The values of ${r_\ast}_\mathrm{min},\:{r_\ast}_\mathrm{max},$ and $N$ are set for each calculation (we have confirmed numerical convergence).

We solve the homogeneous SN equation for the derivation of $A^\mathrm{inc}_{\ell m\omega}$. (Eq.~\eqref{eq:SN_eq}) with
\begin{equation}
\label{eq:boundary_condition}
{r_\ast}_\mathrm{min}(j) = \left\{
\begin{array}{llll}
-40M\:, & j=0.8\\
-50M\:, & j=0.9 \\
-100M\:, & j=0.99 \\
-160M\:, & j=0.999  
\end{array}\;,
\right.
\end{equation}
and 
\begin{equation}
    {r_\ast}_\mathrm{max}(\omega) = 600M + \frac{30M}{\omega}\:\:\:\:\:\mathrm{for\:all}\:j\:.
\end{equation}
$N$ is set to $2\times10^5$.

For the calculation of the integral in Eq.~\eqref{eq:X_infty}, we adopt the same minimum radius as Eq.~\eqref{eq:boundary_condition} and $N=2\times10^3$ for $j=0.8, 0.9, 0.99,$ and $N=5\times10^3$ for $j=0.999$.

\begin{figure*}[t]
\centering
\includegraphics[scale=0.45]{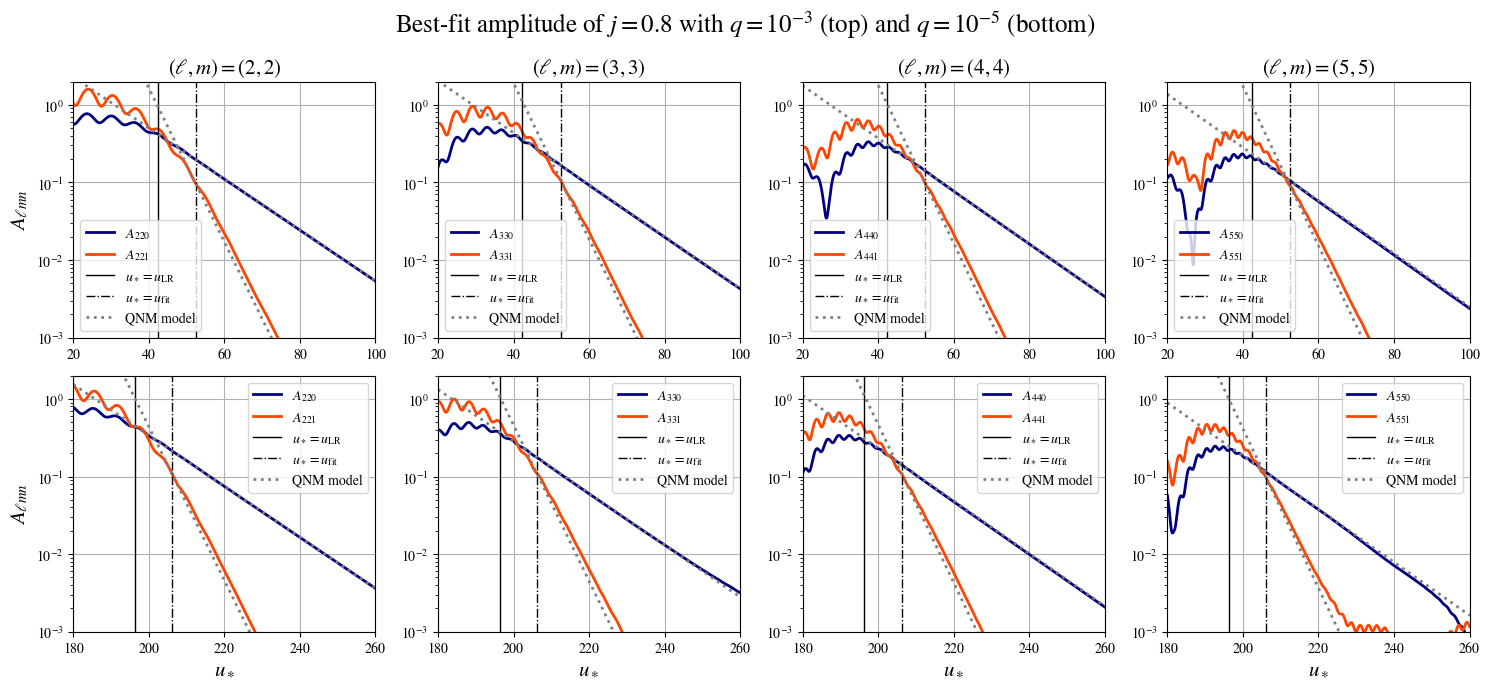}
\caption{Best-fit amplitudes with respect to $u=u_\ast$ for $j=0.8$ and for $(l,m)=(2,2), (3,3), (4,4),$ and $(5,5)$. We set $L_\mathrm{p}=1$ (top) and $L_\mathrm{p}=-1$ (bottom). The best-fit amplitudes $A_{lm0}$ (navy) and $A_{lm1}$ (orange) are shown. The gray dotted lines indicate the amplitudes of each QNM [Eq.~\eqref{eq:qnm_amp}], where the initial value is fixed at the earliest time at which $\mathcal{M}_{lm} \leq 0.001$ for all multipole modes.
} 
\label{fig:amp_08}
\end{figure*}

\begin{figure*}[t]
\centering
\includegraphics[scale=0.45]{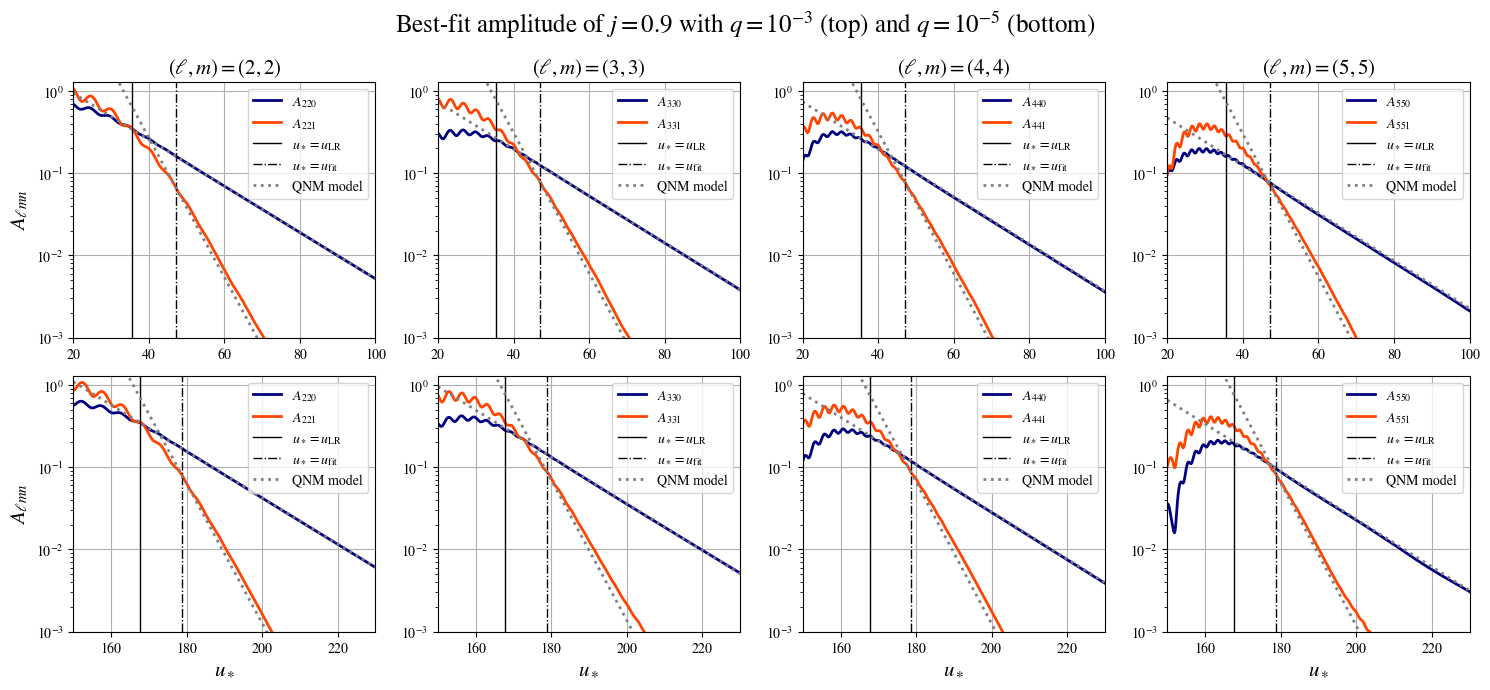}
\caption{Same as Fig.~\ref{fig:amp_08}, but for a primary BH spin of $j=0.9$.
} 
\label{fig:amp_09}
\end{figure*}

\begin{figure*}[t]
\centering
\includegraphics[scale=0.45]{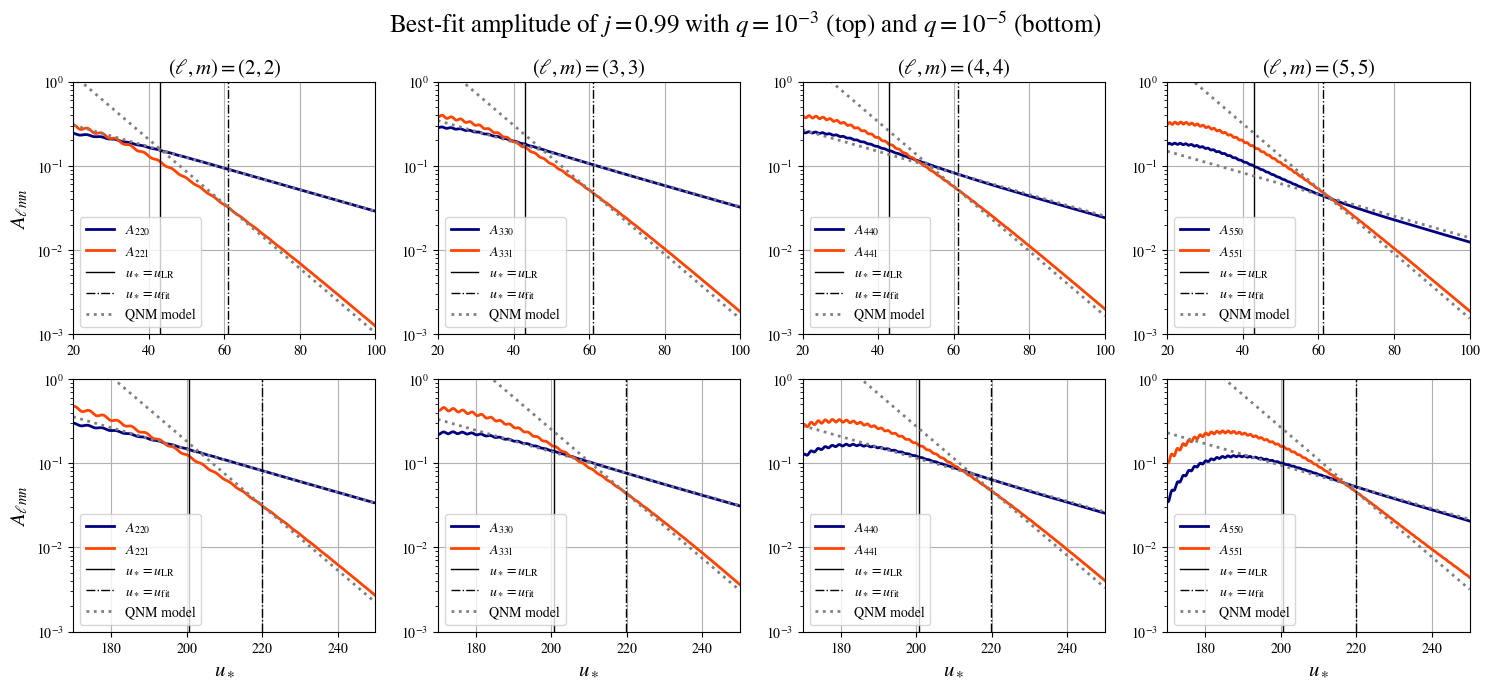}
\caption{Same as Fig.~\ref{fig:amp_08}, but for a primary BH spin of $j=0.99$.
} 
\label{fig:amp_099}
\end{figure*}

\begin{figure*}[t]
\centering
\includegraphics[scale=0.45]{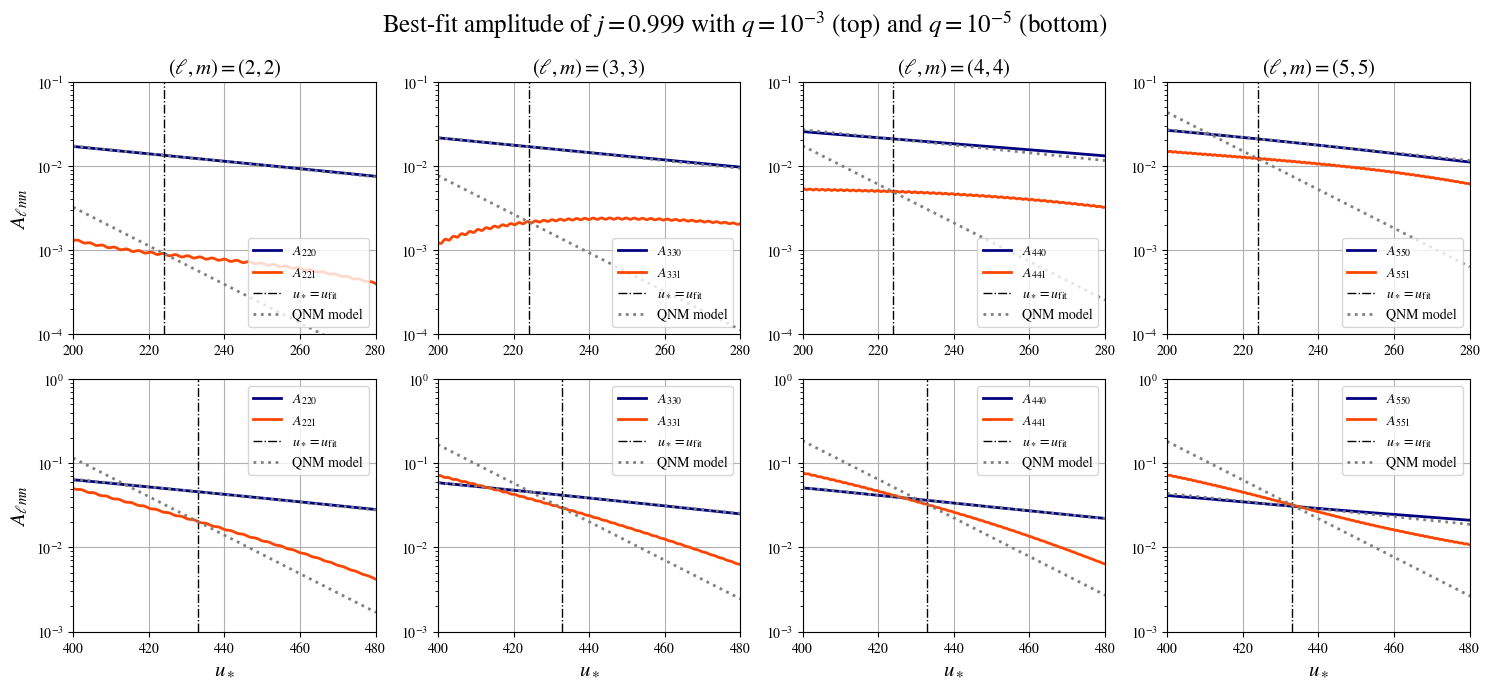}
\caption{Same as Fig.~\ref{fig:amp_08}, but for a primary BH spin of $j=0.999$ and $\mathcal{M}=0.01$ is used to determine $u_\mathrm{fit}$. Amplitude extraction following Eq.~\eqref{eq:qnm_amp} is achieved only for the fundamental mode.
} 
\label{fig:amp_0999}
\end{figure*}

\section{BEHAVIOR OF THE BEST-FIT AMPLITUDE}
\label{app:behavior_amp}
A small $\mathcal{M}_{\ell m}$ does not necessarily ensure that the QNM model works physically. In addition to evaluating $\mathcal{M}_{\ell m}$, we should confirm the behavior of the best-fit amplitudes. If the fits are performed in a self-consistent manner, the extracted amplitude $A_{\ell mn}(u_\ast)$ should follow an exponential decay,
\begin{equation}
\label{eq:qnm_amp}
    A_{\ell mn}(u_\ast) \propto \mathrm{exp}\left[-\frac{u_\ast}{\tau_{\ell mn}}\right]\:,
\end{equation}
where $\tau_{\ell mn}:= \{-\mathrm{Im}(\omega_{\ell mn})\}^{-1}$ is the damping time for the $(\ell,m,n)$ mode. Figures~\ref{fig:amp_08}-\ref{fig:amp_0999} show the best-fit amplitudes of the fundamental mode $(n=0)$ and the first overtone $(n=1)$ for each multipole mode. Except in the case of $j=0.999$, the amplitudes are stably extracted after $u=u_\mathrm{fit}$ until numerical error dominates. In the most highly spinning case, only the fundamental mode's amplitudes are confidently extracted due to the difficulty in distinguishing the two modes caused by the closeness of the QNM frequencies.

\begin{figure}[t]
\centering
\includegraphics[scale=0.55]{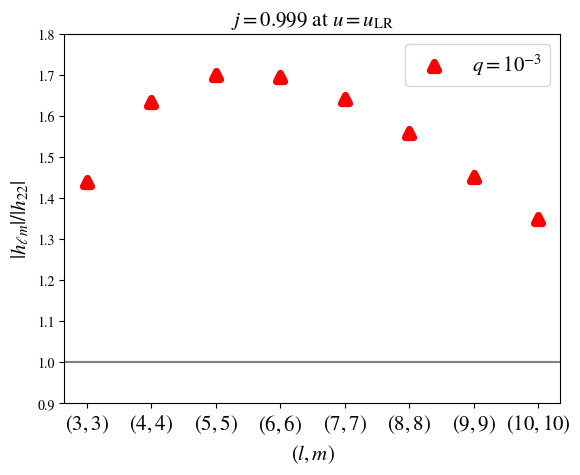}
\caption{Relative amplitudes to the $(2,2)$ mode of the $(\ell, m)=(3,3)$ to $(10, 10)$ modes at $u=u_\mathrm{LR}$ in the intermediate merger with $j=0.999$. 
}
\label{fig:amp_ratio_0999}
\end{figure}

\section{HIGHER HARMONIC QNM EXCITATIONS WITH $ 6 \leq \ell \leq 10 $ IN THE IMR MERGER WITH $j=0.999$}
\label{app:additional_calculation}
We calculate the GW modes with $ 6 \leq \ell \leq 10 $ and evaluate the excitations of higher harmonic QNMs in the intermediate merger $(q=10^{-3})$ with $j=0.999$. The evaluations are done at $u=u_\mathrm{LR}$. In the calculation, we use \texttt{SpinWeightedSpheroidalHarmonics} of the \texttt{Black Hole Perturbation Toolkit}~\cite{BHPToolkit} to derive the angular separation constants for these modes.

Figure~\ref{fig:amp_ratio_0999} depicts the relative amplitudes to the $(2,2)$ mode of from the $(3, 3)$ to the $(10, 10)$ mode in the IMR mergers ($q=10^{-3}$) with $j=0.999$. The result shows the $(5, 5)$ mode gives the maximum, and the $(6, 6)$ mode has a comparable amplitude. Furthermore, this indicates that, as $\ell=m$ increases, the mode amplitude monotonically decreases.

\newpage
\bibliographystyle{apsrev4-2}
\bibliography{reference}

\end{document}